\documentclass[aps,prb,twocolumn,showpacs,floatfix]{revtex4-1}
\usepackage{graphicx}
\usepackage{latexsym}

\usepackage{subfigure}
\usepackage{hyperref}
\usepackage{epsfig}
\usepackage{float}
\usepackage{amsmath}
\usepackage{amssymb}
\usepackage[toc,page]{appendix}

\DeclareRobustCommand{\topleft}{%
  \mathbin{\text{\rotatebox[origin=c]{-90}{$\Bigl [$}}}%
}
\DeclareRobustCommand{\topright}{%
  \mathbin{\text{\rotatebox[origin=c]{90}{$\Bigr ]$}}}%
}
\DeclareRobustCommand{\largetopbracket}{%
  \mathbin{\text{\rotatebox[origin=c]{90}{$\Biggl ]$}}}%
}
\DeclareRobustCommand{\threeleft}{%
  \mathbin{\text{\rotatebox[origin=c]{-90}{$\Bigl [$}}}%
}
\DeclareRobustCommand{\threeright}{%
  \mathbin{\text{\rotatebox[origin=c]{90}{$\Bigr ]$}}}%
}
\begin{document}

\title{Linked cluster expansion of the many-body path integral}
\author{Anish Bhardwaj$^{(1)}$}
\author{ Efstratios Manousakis$^{(1,2)}$}
\affiliation{
$^{(1)}$ Department  of  Physics and National High Magnetic Field Laboratory,
  Florida  State  University,  Tallahassee,  FL  32306-4350,  USA\\
$^{(2)}$Department   of    Physics,   University    of   Athens,
  Panepistimioupolis, Zografos, 157 84 Athens, Greece
}
\date{\today}
\begin{abstract}
We develop an approach of calculating the many-body path integral based on
the linked cluster expansion method. 
First, we derive a linked cluster expansion and we give
the diagrammatic rules for calculating 
the free-energy and the pair distribution
function $g(r)$ as a systematic power series expansion in the particle 
density. We also generalize the hypernetted-chain (HNC) equation for $g(r)$,
known from its application to classical statistical mechanics, 
to a set of quantum HNC equations (QHNC) for the quantum case. 
The calculated $g(r)$ for 
distinguishable particles interacting with a Lennard-Jones potential in various
attempted schemes of approximation 
of the diagrammatic series 
compares very well with the results of path integral Monte Carlo simulation
even for densities as high as the equilibrium density of the strongly
correlated liquid $^4$He. 
Our method is applicable to a wide range of problems of current general interest
and may be extended to the case of identical particles and, in 
particular, to the case of the many-fermion problem. 
\end{abstract}
\maketitle
\section{Introduction}
With the advance of computational power during the last four decades,
Monte Carlo simulations have been very successful in addressing a volume of
issues of classical statistical mechanics. 
As a result, earlier methods based on other approximate computational
techniques, such as the cluster
expansion\cite{Mayer}\cite{goodstein2002states} in conjunction with 
the hypernetted-chain approach (HNC)\cite{hiroike1,hiroike2,morita1,morita2,morita3,morita4,morita5} have become much less popular and
are rather rarely used.
Furthermore, the Monte Carlo method has been extended to carry out 
simulations of quantum many-body systems. It has 
been successfully developed to treat accurately a variety of problems
including, quantum-spin systems, the polaron 
problem\cite{PhysRevLett.81.2514,PhysRevB.62.6317}, and bosonic 
systems in the regime of
high particle-density and as strongly correlated as
liquid $^4$He\cite{RevModPhys.67.279}.
There are problems, however, such as the many-fermion 
systems and frustrated quantum spin systems, where there is no useful a priori 
positive-definite quantity to be used as the sampling probability in 
the Markov process. There are many useful ideas on how to treat these problems
within some approximation scheme, such as, the fixed node approximation\cite{RevModPhys.84.1607} or the constrained path Monte Carlo\cite{PhysRevB.55.7464}, 
however, they are limited in the regime of their applicability.

In the present paper we develop a cluster expansion technique to
treat the quantum many-body Feynman path-integral\cite{Hibbs} at 
finite-temperature\cite{feynman1998statistical}. The expansion 
can be regarded as a systematic expansion in the number of particles involved in the
diagrams retained, i.e., keeping up to $n$-body clusters.
Alternatively, this  can be viewed as a systematic formal expansion in the 
particle density $\rho$.
We tested the method by comparing our results for the pair distribution function $g(r)$ for distinguishable particles
interacting with the Lennard-Jones interaction  to the results obtained
by the path integral Monte Carlo (PIMC) method.
Our results for $g(r)$ in a wide distance-range 
obtained by keeping just up to three-body clusters are in good agreement with the results of
the PIMC method even at  $^4$He densities down to moderately low temperature.
Furthermore, we generalize the
HNC resummation approach to the quantum case (QHNC). We show that
our results for distinguishable particles described by the Lennard-Jones
interaction are in good agreement with the results obtained by the 
PIMC method for the same system. In particular, 
we are encouraged by the fact that
the results for $g(r)$ are accurate for the high-density regime applicable to helium where short-range correlations are strong.

This method is generalizable to the case of identical particles and, therefore,
it has the potential to be useful as an alternative approach to the treatment of
fermionic systems or frustrated quantum-spin systems.
While the approach of keeping up to $n$-body clusters and the QHNC approach are both approximations, 
they may provide  alternative approaches to  complex problems from a different angle.
For example, while at high density the electron gas\cite{PhysRevLett.45.566}  is accurately described by the random phase
approximation \cite{fetter2003quantum,PhysRev.92.609},
at low-density, there is no exact treatment.  
Our method, which becomes asymptotically exact in the low-density limit,
could provide the still missing accurate approach for the electron gas at low densities. 
Furthermore, the approach developed in the present work is a general method to treat
the quantum many-particle problem.
Therefore, it can be applied to other systems in diverse areas of physics,
such as, systems of trapped ultra-cold atoms, and possibly 
to the many-nucleon problem\cite{RevModPhys.51.821},
i.e., the hypothetical infinite nuclear matter and neutron stars.

The paper is organized as follows. In Sec.~\ref{path-integral} we discuss
the many-body path integral and we cast it in a form useful for
the application of our method. In Sec.~\ref{partition-function} we develop
the cluster expansion of the path integral that describes the
quantum mechanical partition function for a  system of distinguishable particles. 
We give the diagrammatic rules for a systematic
inclusion of all the diagrams order by order in the density.
In Sec.~\ref{sec:free-energy} we discuss the cluster expansion of the
free-energy and in Sec.~\ref{distribution-function}
that of the pair distribution function.
In Sec.~\ref{summation} we present out results of the method viewed as
a systematic power series expansion in the density. In addition, we
derive the quantum hypernetted-chain (QHNC) equations and we give
our results for the Lennard-Jones system for densities near the equilibrium
density of liquid $^4$He. Last, in Sec.~\ref{conclusions} we present our
conclusions.

\section{The many-body path integral}
\label{path-integral}
To build the path integral cluster expansion formalism we will follow the same procedure that is used in the case of 
classical cluster expansion. While our goal in this paper is to study the simpler case of distinguishable
particles, we first begin by writing down the partition function \cite{feynman1998statistical} for a system of 
$N$ interacting identical particles:
\begin{eqnarray}
Z&=&\frac{1}{N!}\underset{P}{\sum}(\pm1)^{[P]}\int  d^{3}\vec r_{1}d^{3}\vec r_{2}....d^{3}\vec r_{N} \nonumber \\
& &\langle\vec{r}_{1},\vec{r}_{2},...,\vec{r}_{N} 
\mid e^{-\beta\hat{H}}\mid\vec{r}_{P1},\vec{r}_{P2},...,\vec{r}_{PN}\rangle.
\end{eqnarray}
Here, the summation over $P$ means a summation over all permutations of particles and the notation $[P]$ denotes the order of the permutation\cite{feynman1998statistical}.
In the case of bosons all permutations contribute with a positive sign.
In the case of fermions, $(-1)^{[P]}$ is +1 or -1 depending on whether the permutation is even or odd. In the positions $(\vec r_{P1},\vec r_{P2}, ..., \vec r_{PN})$ the indices  $P1, P2, ..., PN$ are the particle indices
after permutation $P$. 
In this paper we will deal with the general case of an interacting
Hamiltonian $\hat{H}$ of the following form:
\begin{eqnarray}
\hat{H}=-\frac{\hbar^2}{2m}\overset{N}{\underset{i}{\sum}}\nabla_i^2+\underset{i<j}{\sum}v(r_{ij}).
\end{eqnarray}
Next, we divide the imaginary-time interval $[0,\theta]$
($\theta\equiv \hbar\beta$) into $M$ slices
of size $\delta \tau = \hbar\beta/M$ to write $e^{-\beta \hat H} =
e^{-\delta \tau \hat H/\hbar} e^{-\delta \tau\hat H/\hbar } ...e^{-\delta \tau\hat H/\hbar } $.
As usual we insert in between each pair of these operators the unit operator
expressed as a sum over the complete set of many-body position eigenstates and
by using the Trotter approximation we can write the partition function as:
\begin{eqnarray}
Z&=&\underset{P}{\sum}\frac{(\pm1)^{[P]}}{N!}\int_{\vec r_i(\theta) = \vec r_{Pi}(0)}\prod_{n=1}^N\prod_{k=0}^{M-1}\frac{d^{3}\vec{r}_{n}^{(k)}}{\lambda^3_{\delta\tau}}e^{-S_E},
\label{part-M}
\end{eqnarray} 
where,
\begin{eqnarray}
S_E &=&\frac{\delta\tau}{\hbar}\overset{M-1}{\underset{k=0}{\sum}}\underset{i<j}{\sum}\Bigl [\frac{m}{2{(\delta\tau)}^2}r^2_{i}(kk+1)
  + v(r_{ij}^{(k)})\Bigr ], \label{E_action}
\\
\lambda_{\tau} &\equiv& {(2\pi\hbar\tau/m)^{1/2}},
\end{eqnarray}
and $r^{(k)}_{ij}= | \vec r^{(k)}_{i}-\vec r^{(k)}_{j}|$ and $r_i(kl)=|\vec{r}_i^{(k)}-\vec{r}_i^{(l)}|$.
The constraint ${\vec r_i(\theta) = \vec r_{Pi}(0)}$ on the path integral
(where $\theta = \hbar \beta$) means that the sum is over all possible $N$-particle paths which start
at positions $(\vec r_1(0),\vec r_2(0), ..., \vec r_N(0))$ at imaginary-time $\tau=0$ and after the ``lapse'' of an  imaginary-time interval of 
$\beta \hbar$ they end up at positions $(\vec r_1(\theta)= \vec r_{P1}(0),
\vec r_2(\theta) = \vec r_{P2}(0), ..., \vec r_N(\theta) = \vec r_{PN}(0))$
where $P1, P2, ..., PN$ are the particle indices after permutation $P$.

By taking the limit $M\rightarrow\infty$, one can obtain the well-known Feynman's path-integral expression:   
\begin{equation} 
Z=\frac{1}{N!}\underset{P}{\sum}(-1)^{[P]}\int_{\vec r_i(\theta) = \vec r_{Pi}(0)} {\cal D}\vec r_{1}{\cal D} \vec r_{2} ... {\cal D}\vec r_{N}e^{-S_E},
\end{equation}
The Euclidean action in the above path integral is given by 
\begin{eqnarray}
S_E=\int_0^{\hbar\beta}(\sum_{i=1}^{N}\frac{1}{2}m\dot{\vec{r}}_{i}^{2}+\sum_{i<j}v(r_{ij}(\tau)))d\tau,
\end{eqnarray}
where $r_{ij}(\tau) = |\vec{r_{i}}(\tau)-\vec{r}_{j}(\tau)|$.

However, the Feynman's path-integral expression is only symbolic and, for all practical purposes, we make use of the expression given by Eq.~\ref{part-M}.
As discussed in the abstract and in the introduction of this paper, in
this  paper  we concentrate  our  attention  to  the simpler  case  of
distinguishable  particles,  which   corresponds  to  considering  the
identity permutation only.

\section{Cluster expansion of partition function}
\label{partition-function}
In this  paper we concentrate  only on the identity  permutation which
corresponds  to a  system  of  interacting distinguishable  particles.
This will allow us to test the  method by comparison of our results to
PIMC which is accurate in this case because of the absence of the sign
problem.   Since our  method  is  diagrammatic in  nature,  it can  be
extended   to   include   diagrams  which   correspond   to   particle
permutations. Therefore, if  we can demonstrate that  the method works
for distinguishable  particles, it would  be a promising sign  for the
applicability of the  method to the more-complex  problem of identical
particles and in particular the problem of Fermions where QMC fails to
address  it  in   an  exact  way.  While  we  consider   the  case  of
distinguishable particles,  as we describe  our method in  the present
paper,  when appropriate,  we  address the  generalizations needed  in
order to include the permutations.

Now, we concentrate on the dimensionless ratio $Z\over Z_0$ 
where  $Z_0$ is the non-interacting partition function, which can be obtained from 
Eq.~\ref{part-M} by  using the free-action, obtained from Eq.~\ref{E_action} with $v(r^{(k)}_{ij})=0$,
and carrying out the Gaussian integrals:
\begin{eqnarray}
  Z_0=\frac{1}{N!}\Bigl (\frac{V}{V_{\theta}} \Bigr )^N, \hskip 0.2 in
  V_{\theta}=\lambda^3_{\theta},
\label{F(beta)}
\end{eqnarray} 
and since $\lambda_{\theta}$ is the de Broglie thermal wavelength,
$V_{\theta}$ is the de Broglie thermal volume. Thus,
\begin{eqnarray}
\frac{Z}{Z_0}= \Bigl (\frac{V_{\theta}}{V} \Bigr )^N Z.
\label{ratio} 
\end{eqnarray}

Next, we start the cluster expansion by defining the
function $h_{ij}^{(k)}$  as follows:
\begin{eqnarray}
e^{-\frac{\delta\tau}{\hbar}v(r_{ij}^{(k)}) } &\equiv&
1+h_{ij}^{(k)}, \label{h-factor}
\end {eqnarray}
and the function  $L_i(kl)$ as follows:
\begin{eqnarray}
L_i(kl)&=&\frac{1}{\lambda^3_{\tau_{kl}}}
e^{-\pi \frac{r^2_i(kl)}{\lambda^2_{\tau_{kl}}}}, \label{world-line}\\
\tau_{kl} &=& |k-l| \delta \tau.
\end {eqnarray}
Using these definitions the partition function can be written as
\begin{eqnarray}
{ Z \over {Z_0}} &=& \Bigl (\frac{V_{\theta}}{V} \Bigr )^N \int 
\prod_{n=1}^N\prod_{k=0}^{M-1}d^{3}r_n^{(k)} L_n(kk+1) \nonumber \\
&\times&\prod_{l=0}^{M-1}\prod_{i<j}(1+h_{ij}^{(l)}).
\label{part1}
\end{eqnarray} 

\begin{figure}
    \vskip 0.3 in \begin{center}
        \subfigure[]{
           \includegraphics[scale=0.35]{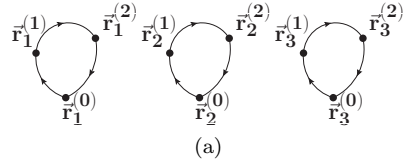}
           \label{fig:2a}
        }
\hskip 0.2 in
        \subfigure[]{
            \includegraphics[scale=0.35]{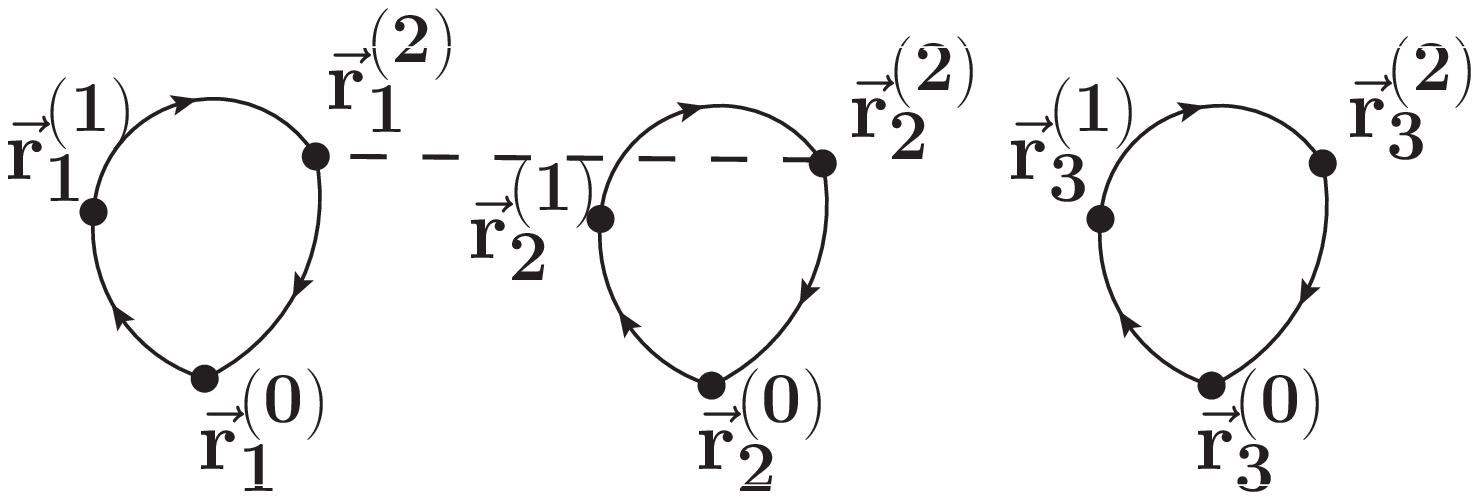}
            \label{fig:2b}
        }

\vskip 0.2 in
        \subfigure[]{
            \includegraphics[scale=0.35]{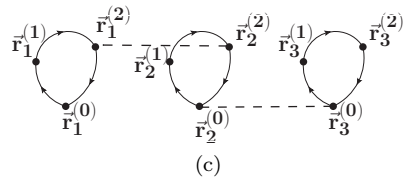}
            \label{fig:2c}
        }
\hskip 0.2 in
        \subfigure[]{
            \includegraphics[scale=0.35]{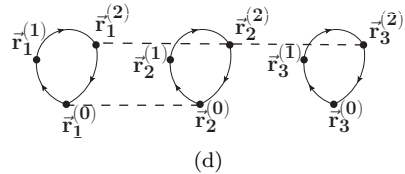}
            \label{fig:2d}
        }

    \end{center}
\caption{\label{fig:fig2}Few diagrams that appear in the expression for $Z$ when $N=3$ and $M=3$.}
\vskip 0.2 in \end{figure}

The expression in Eq.~\ref{part1}  consists of products of $(1+h_{ij}^{(l))})$,
which can be written out as follows:
\begin{eqnarray}
\prod_{l=0}^{M-1}\prod_{i<j}(1+h_{ij}^{(l)})=1+\sum_{l=0}^{M-1}\sum_{i<j}h_{ij}^{(l)}
+ ...
\label{h-products}
\end{eqnarray}
where we have omitted terms containing two or more $h$ factors.
After substituting the expanded product in Eq.~\ref{ratio} we 
obtain a sum of integrals. We can keep track of terms by representing
each of these integral terms by diagrams. 
As an example consider the case of 3 ($N=3$) particles with 3 time instants 
($M=3$). In this case, the first few terms have been diagrammatically 
represented in Fig.~\ref{fig:fig2}.

We use the following convention for representing such terms:
\begin{enumerate}
\item  The positions of the particles at the instants of time which enter in the
integral are denoted by solid
circles. An integration over these positions is implied. 

\item $h_{ij}^{(k)}$ is denoted by a dashed line labeled by the index $(k)$ (which represents interaction between particles at the same instant of time $k$) 
connecting points $i$ and $j$. 
\item The $L$ function defined by Eq.~\ref{world-line}
is denoted by a solid line connecting points $\vec{r}_i^{(k)}$ and $\vec{r}_i^{(l)}$. 

\item Every particle is associated with its own world-line, which is made up of products of $L$ functions.
 The world-lines of a pair of particles at a given instant of time can  
be either disconnected
(i.e., there is a factor of unity)  or they can be connected by 
a dashed-line. 

\item The world-line starts at $\vec{r}_{1}^{(0)}$ and connects back to $\vec{r}_{1}^{(0)}$ because of the boundary condition $\vec{r}_{i}^{(0)}=\vec{r}_{i}^{(M)}$.
Therefore, a world-line forms a loop formed by the particle positions at 
all instants of time.

\end{enumerate}

In the case of identical particles,  there are diagrams in which $\vec
r_i^{(M)}  =  \vec r_{Pi}^{(0)}$,  i.e.,  the  particle positions  are
exchanged at imaginary-time slice $M$. In the present paper, we do not
deal with the contribution of such diagrams.

\begin{figure}
    \vskip       0.3       in       \begin{center}       \subfigure[]{
        \includegraphics[scale=0.35]{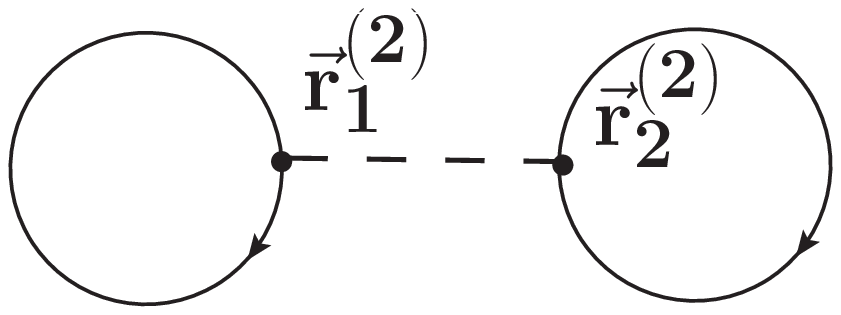}
            \label{fig:1hline}
        }
\hskip 0.2 in
        \subfigure[]{
            \includegraphics[scale=0.35]{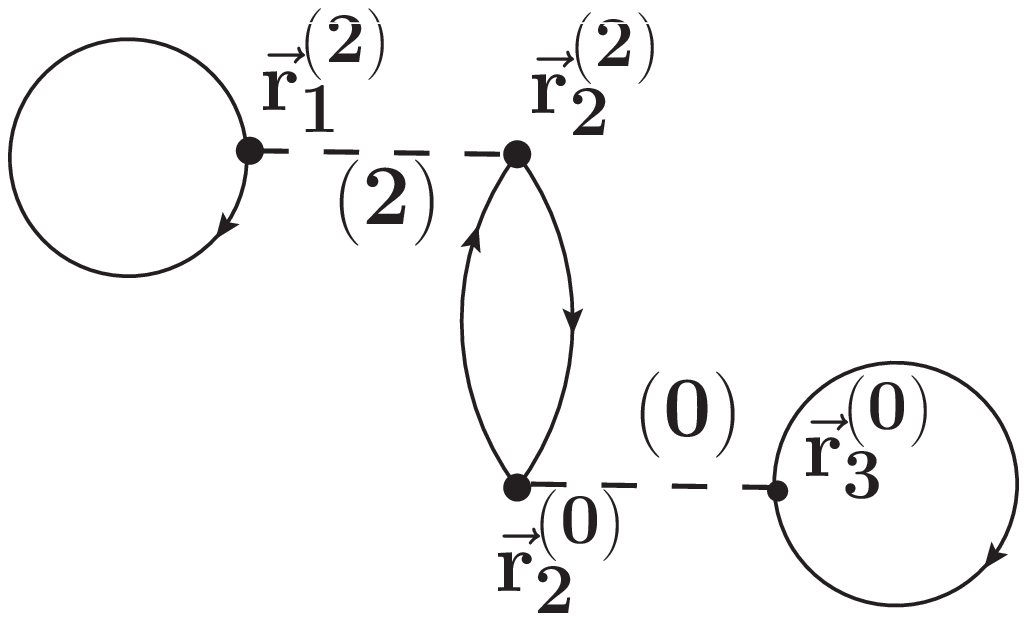}
            \label{fig:2hline}
        }
\end{center}
\caption{\label{fig:1h} (a) and (b) show respectively simplified versions of the 
  diagrams shown  in Fig.~\ref{fig:2b} and in Fig.~\ref{fig:2c}
  after integration over the coordinates
  of those points which 
  are not connected through $h$-lines to any other point.
  } 
\vskip 0.2 in \end{figure}

If the world-line  of a given particle has solid  points which are not
connected to any other point  through dashed-lines then it is possible
to perform the  integration over those variables exactly  by using the
following identity:
\begin{eqnarray}
L_i(kl)=\int d^3\vec{r}_i^{(m)}L_i(km) L_i(ml).
\end{eqnarray}
This result  can be interpreted such  that the world-line now  makes a
straight connection of the particle  coordinate at the initial instant
of time,  i.e., $\vec r_i^{(k)}$  with the particle coordinate  at the
instant  of  time  $\l$.    This  explicit  integration  removes  such
intermediate points  in our  diagrams which are  not connected  by any
dashed-line.     As   an    example,   consider    the   diagram    in
Fig.~\ref{fig:2a}.   Because  of the  absence  of  dashed-lines it  is
possible to perform  the integration over all three  variables and the
diagram equals to  unity; Similarly, the term  in Fig.~\ref{fig:2b} is
obtained when  the product is such  that we have only  one dashed-line
(in this  case $h_{12}^{(1)}$); here,  particle 1 and 2  are connected
but  particle  3 is  disconnected.   Such  a  term has  the  following
integral form:
\begin{eqnarray}
\Bigl                    ({{V_{\theta}}\over                   V}\Bigr
)^{3}\int\underset{i=1}{\overset{3}{\prod}}\underset{k=0}{\overset{2}{\prod}}d^3\vec{r}_{i}^{(k)}L_i(kk+1)h_{12}^{(2)}.
\end{eqnarray}
The  coordinates  of  particle  3  at all  instants  of  time  can  be
integrated out  and, thus,  the diagram  corresponds to  the following
expression:
\begin{eqnarray}
\Bigl             ({{V_{\theta}}\over             V}             \Bigr
)^{2}\int\underset{i=1}{\overset{2}{\prod}}\underset{k=0}{\overset{2}{\prod}}d^3\vec{r}_{i}^{(k)}L_i(kk+1)h_{12}^{(2)}.
\end{eqnarray}
We  can further  simplify  the  above expression  by  noting that  the
integration      over       $\vec{r}_1^{(0)}$,      $\vec{r}_2^{(0)}$,
$\vec{r}_1^{(1)}$, and $\vec{r}_2^{(1)}$ can be performed as these are
not connected through a dashed-line.   The simplification leads to the
following term:
\begin{eqnarray}
&\Bigl         ({{V_{\theta}}\over          V}\Bigr         )^{2}&\int
  d^3\vec{r}_{1}^{(2)}d^3\vec{r}_{2}^{(2)}L_1(0M)L_2(0M)h_{12}^{(2)}
  \nonumber      \\      &=&       {1      \over      {V^2}}      \int
  d^3\vec{r}_{1}^{(2)}d^3\vec{r}_{2}^{(2)} h_{12}^{(2)}.
\label{eq:1h}
\end{eqnarray}
Such a term has been represented in Fig.~\ref{fig:1hline}. Each of the two $L$
functions corresponding  to the world-lines  of particle  1 and 2 which
begin and go back to the same position at zeroth time-slice
(because $\vec r^{(M)}_i=\vec r^{(0)}_i$) yield the constant factor
 $L_i(0M)=1/V_{\theta}$.
In a similar way, the diagram in Fig.~\ref{fig:2c}, which is an example of a 
connected cluster, can be represented as shown in Fig.~\ref{fig:2hline} 
with a contribution given by:
\begin{eqnarray}
 {{{V_{\theta}}} \over {V^3}} \int d^3\vec{r}_{1}^{(2)}d^3\vec{r}_{2}^{(2)} d^3\vec{r}_{2}^{(0)}&&d^3\vec{r}_{3}^{(0)} L_2(02) L_2(20)
h_{12}^{(2)}\times\nonumber \\
&&h_{23}^{(0)}.
\end{eqnarray}

\subsection{Diagrammatic rules}
Now let us consider the expansion of $Z/Z_0$ for a very large number of particles $N$. In such an expansion we can
still have terms which have the same expression as, for example the diagram of Fig.~\ref{fig:1hline} (Eq.~\ref{eq:1h}),
in which the coordinates of all the other particles except 1 and 2 have been integrated out because they were not 
connected to any other particle by an $h$-line. In addition, the exact same two-body contribution arises when the
labels of particles 1 and 2 are interchanged with any of the other $N-2$ particles. Therefore, 
when we draw a diagram such as  the diagram of Fig.~\ref{fig:1hline} we imply that we include all the $N(N-1)/2$ diagrams
which correspond to those obtained from replacing 1 and 2 with any other pair of particles. Therefore, the contribution of this
diagram is going to be:
\begin{eqnarray}
{1 \over 2} (V_{\theta} \rho)^{2}\int d^3\vec{r}_{1}^{(2)}d^3\vec{r}_{2}^{(2)}L_1(0M)L_2(0M)h_{12}^{(2)},
\label{eq:1h-p}
\end{eqnarray}
where we have used the fact that as $N\to \infty$, $N(N-1)/V^2 \to \rho^2$.
To summarize, the expansion of $Z/Z_0$ can be obtained as a summation of
terms which correspond to 
\begin{eqnarray}
{Z \over {Z_0}} = 1 + \sum_{n=2}^{\infty} \sum_{\alpha} D^{(\alpha)}_n,
\label{eq:zoverz0}
\end{eqnarray}
when $D^{(\alpha)}_n$ stands for any $n$-body diagram. Here, $n$ is the number 
of particles involved in the diagram and $\alpha$ labels the various
$n$-body diagrams. An $n$-body diagram is a diagram with  $n$ particles 
connected to each other through dashed-lines and each of the remaining 
$N-n$ particles are not connected to any of the other particles.
In this case, the coordinates
of the latter $N-n$ particles drop out. 
To find all $D^{(\alpha)}_n$ we need to draw all topologically
distinct $n$-particle diagrams by following the following rules:

\begin{enumerate}
\item  Particle positions  are denoted by 
solid dots labeled $\vec r_{i}^{(k)}$ and they stand 
for an integration over 
the coordinate of the $i$th particle at the $k$-th instant of time.

\item We need to select the $n$ world-lines for each one of the
$n$ particles. Each world-line starts at time $\tau=0$ and ends
at time $\tau = \hbar \beta$ and it is made out of connected   
solid-lines which correspond to $L$ functions (see Eq.~\ref{world-line}) 
 which connect particle coordinates at intermediate instants
of time. The integrations over particle coordinates at intermediate instants of time 
are allowed unless they are connected to the world-line of at 
least one other particle by a dashed-line at the same instant of time.

\item We choose to connect pairs of particle positions $\vec r^{(k)}_i$, and
$\vec r^{(k)}_j$ at the same instant of time $k$ by dashed-lines. Each such
 dashed-line labeled as $(k)$ connecting particles $i$ and $j$ 
gives rise to a factor $h_{ij}^{(k)}$.

\item For every world-line we need to multiply the contribution of the diagram by a {\it dimensionless} factor of ${\rho V_{\theta}}$.

\item The contribution of a diagram is divided by a factor of $S$, the symmetry factor of the diagram. In the case
of the diagram of Fig.~\ref{fig:1hline} the factor 1/2 in Eq.~\ref{eq:1h-p} is due to a symmetry factor of 2 due to 
the fact that by interchanging the two points 1 and 2 the contribution of the diagram remains the same.

\end{enumerate}

\subsection{Connected and disconnected diagrams}
\begin{figure}
\vskip 0.3 in \begin{center}
\includegraphics[scale=0.35]{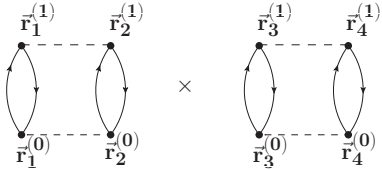}
\end{center}
\caption{\label{fig:discon} Example of a disconnected term (see Eq.~\ref{discon}).} 
\vskip 0.2 in \end{figure}

An $n$-body diagram is considered a {\it connected 
diagram} when each of the $n$ particles is connected to at least one
of the other $n$-particles in the diagram.
All the diagrams that appear in Fig.~\ref{fig:fig2} are examples of connected 
diagrams.   On the other hand, we have a disconnected  
 $n$-body diagram when it is formed out of subsets of particles
 which are connected in such a way 
that particles of any given set are connected to each other and they remain 
disconnected from any particle not belonging to the given set.
A simple example (with $M=2$) of a disconnected term 
is shown in Fig.~\ref{fig:discon} and it is given by:
\begin{eqnarray}
\Bigl ({{V_{\theta}}\over V}\Bigr )^{4}&&\int\underset{i=1}{\overset{2}{\prod}}\underset{k=0}{\overset{1}{\prod}}d^3\vec{r}_{i}^{(k)}L_i(kk+1)h_{12}^{(0)}h_{12}^{(1)}\times \nonumber\\
&&\int\underset{i=3}{\overset{4}{\prod}}\underset{k=0}{\overset{1}{\prod}}d^3\vec{r}_{i}^{(k)}L_i(kk+1)h_{34}^{(0)}h_{34}^{(1)}.
\label{discon}
\end{eqnarray}
\subsection{Factorizable diagrams}
\begin{figure}
    \vskip 0.3 in \begin{center}
        \subfigure[]{
           \includegraphics[scale=0.35]{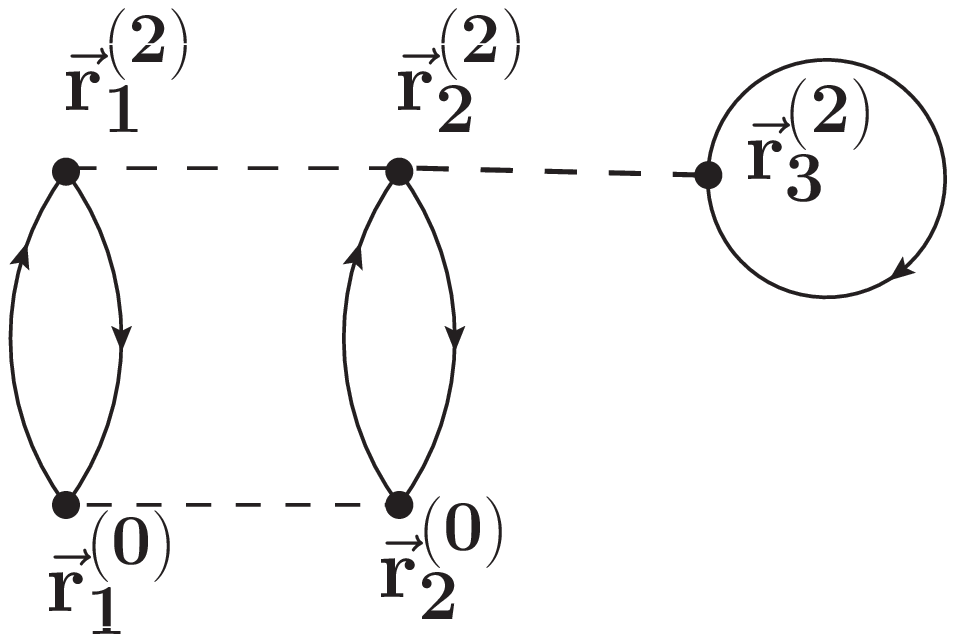}
           \label{fig:5a}
        }
\hskip 0.2 in
        \subfigure[]{
            \includegraphics[scale=0.35]{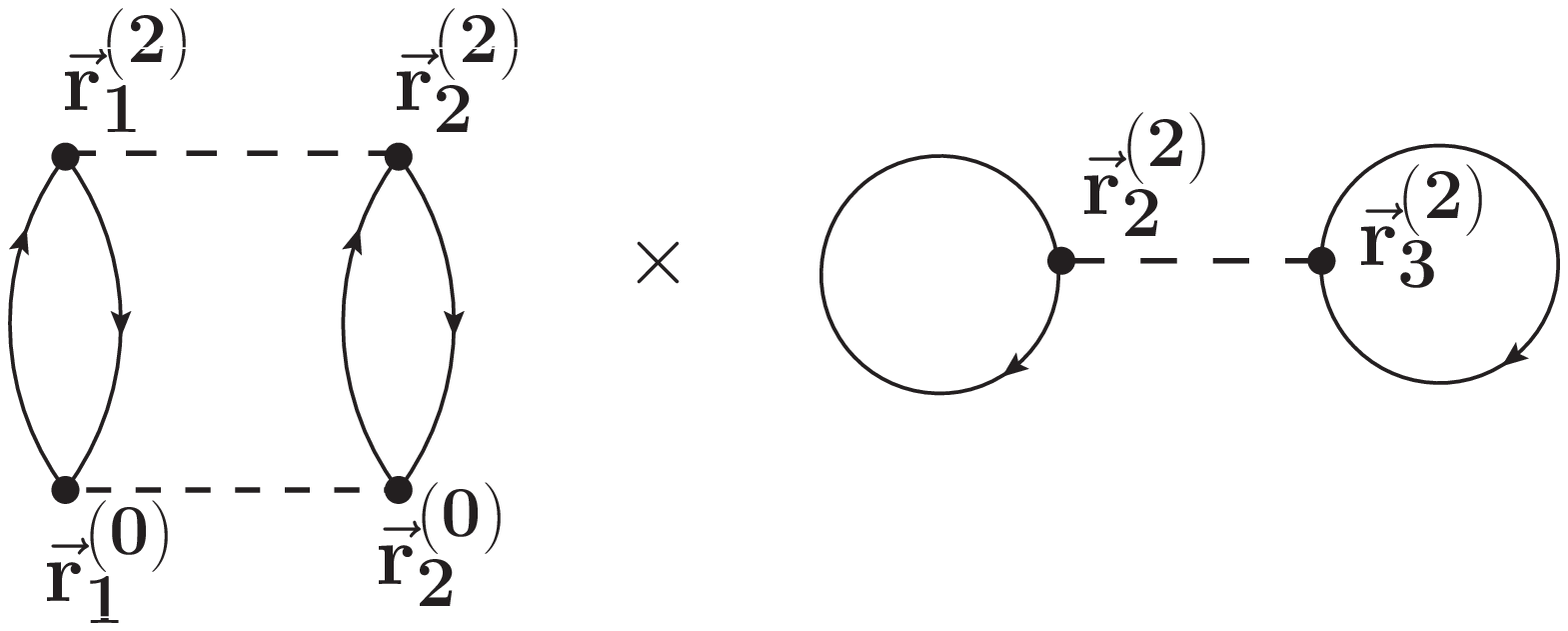}
            \label{fig:5b}
        }
\caption{\label{fig:factorized} An example of a factorizable diagram.}
\end{center}
\vskip 0.2 in \end{figure}

There are  diagrams which can  be factorized  into products of  two or
more  different  diagrams.   Consider,  for example,  the  diagram  of
Fig.~\ref{fig:5a}.   This diagram  is factorizable  at the  node $\vec
r_2^{(2)}$ at which point the world-line  of particle 2 connects via a
dynamical $h$-line to the world line  of particle 3.  This diagram can
be   written  as   a  product   of   two  parts   as  illustrated   in
Fig.~\ref{fig:5b}.  In order  for a  diagram to  be factorizable,  two
parts of  the diagram should  be connected only  at a node,  namely, a
point through which one has to go through when traveling from one part
of the diagram to the other part using $h$-lines or $L$-lines.

\section{Free Energy}
\label{sec:free-energy}
In the expansion of $Z/Z_0$  it is possible to re-group the various diagrams 
in such a way that it can be written as a sum of connected and disconnected clusters.  
We will use the notation $[i]$ to denote the sum of all connected clusters.  It is  easy to see that any disconnected cluster can be written in terms of product of connected clusters. We can write the sum of all the diagrams
contributing to $Z/Z_0$  as follows\cite{RevModPhys.51.821}:
\begin{eqnarray}
\frac{Z}{Z_0}=1 + [i] + \frac{1}{2} [i][j] + \frac{1}{3!}[i][j][k] + ...
\label{cummulant1}
\end{eqnarray}
where $[i]$ denotes the $i$th connected piece and 
a summation over $i$,$j$,$k$, ... is implied. In addition, the
notation $[i][j]$ means a disconnected diagram made out of two parts
where no-common particle exists.
The second term is the sum of all connected diagrams. The third term is the sum of all the disconnected diagrams which are products of just two connected pieces.
The factor of 1/2 is present to avoid double counting of terms in which 
 $[i]$ and $[j]$ are interchanged. 
Similarly, we have a factor of $1\over{3!}$ in the fourth term which is the
sum of all disconnected diagrams made out of three connected pieces.

The free-energy $F(N,T,V)$ of  the system is obtained as
\begin{eqnarray}
 F(N,T,V) = -k_B T N \ln \Bigl ( V/V_{\theta} \Bigr ) - k_B T \ln \Bigl (\frac{Z}{Z_0}\Bigr ).
\end{eqnarray}
Therefore, the corrections to the ideal gas Free-energy is given by
\begin{eqnarray}
-\beta \delta F(N,T,V) =  \ln \Bigl (\frac{Z}{Z_0}\Bigr ).
\end{eqnarray}
Using the expression  in Eq.~\ref{cummulant1} and the Taylor expansion 
of $ln(1+x)$ 
with $x=[i] + \frac{1}{2} [i][j] + \frac{1}{3!}[i][j][k] + ... $, we obtain
\begin{eqnarray}
\ln \Bigl (\frac{Z}{Z_0}\Bigr ) &=& [i] + { 1 \over 2} [i][j] + {1 \over 6} [i][j][k] -  {1 \over 2} [i]\times[j]  \nonumber \\
&-& {1 \over 2} [i]\times[j][k] +  
{1 \over 3} [i]\times[j]\times[k] + ...,
\label{cummulant2}
\end{eqnarray}
and the above expression is correct up to terms which contain less than 4 disconnected clusters.
Using the notation of Ref.~\onlinecite{RevModPhys.51.821}:
\begin{eqnarray}
[i]\times [j] = [i][j] + [{\overset{\topleft} {i][j}}] + 
[{\overset{\topleft} {\overset{\topleft} {i][j}}}] + ...,
\label{overhead}
\end{eqnarray}
where one overhead bar means that the two clusters share one common particle and
two such bars imply that they share two particles. Each diagram in $[i]$ is
of the order of $N$ and the diagrams contributing to $[i][j]$ are of order of 
$N^2$, etc.  As a result, the term  $[{\overset{\topleft} {i][j}}]$ is of order $N$ and
the term $[{\overset{\topleft} {\overset{\topleft} {i][j}}}]$ is of order unity. In general
when there is such an overhead bar, it reduces the order of the contribution by a power of $N$. Substituting  Eq.~\ref{overhead} in Eq.~\ref{cummulant2}
we obtain the following terms up to order $N$:
\begin{eqnarray}
-\beta \delta F(N,T,V) &= & [i] - {1 \over 2} [{\overset{\topleft} {i][j}}] +
{1 \over 2} [\overset{\largetopbracket}{\overset{\topleft} {i][j} ][k]} +\nonumber \\
&&{1 \over 3} \overset{\topleft\topright} {[i][j][k]} + ..., 
\end{eqnarray}
where $\overset{\topleft\topright} {[i][j][k]}$, means that the three
pieces share the same particle. Also,here the ellipses stand for terms containing products of more than 
three disconnected parts with common particles. Notice that each term is of order $N$: A disconnected piece is
of order $N$ and every overhead line removes a factor of $N$. As long as each term has just one overhead line less than
the number of its disconnected pieces, the contribution of the term is of order $N$.

\begin{figure}
\vskip 0.3 in \begin{center}
\includegraphics[scale=0.35]{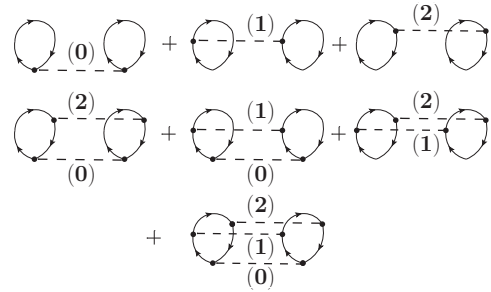}
\end{center}
\caption{\label{free-energy-first-order} 
Diagrams contributing to the free-energy which are first order in $\rho$. }
\vskip 0.2 in \end{figure}

We can simplify the above expression by making use of factorizability. We use the notation $[a]$ to denote non-factorizable type diagram. With this notation $[i]$ can be written as a sum of connected diagrams with all possible factorizable pieces.
\begin{eqnarray}
[i]=[a] + {1\over 2}[{\overset{\topleft}{a\hskip 0.1 in b}}] + ... .
\end{eqnarray}
So that,
\begin{eqnarray}
-\beta \delta F(N,T,V)=[a] + \frac{1}{2}([\overset{\topleft}{a\hskip 0.1 inb}]-[\overset{\topleft}{a][b}])+... .
\end{eqnarray}
Fig.~\ref{free-energy-first-order} shows diagrams contributing to the free-energy which are first order in $\rho$. 

\section{Cluster expansion of distribution function}
\label{distribution-function}

We now develop a diagrammatic expansion for the pair distribution function. We consider the distinguishable particle case by taking only the identity permutation. The pair distribution function for an isotropic translationally invariant 
system takes the following form in our notation:
\begin{eqnarray} 
g(r)&=& {{\cal N} \over {\cal D}}, \hskip 0.2 in  
{\cal D} \equiv {Z \over {Z_0}}, \\
{\cal N} &\equiv& {{V} \over {Z_0}}
\int \underset{i=1}{\overset{N}{\prod}}\underset{k=0}{\overset{M-1}{\prod}}
\frac{d^{3}r_i^{(k)}} {\Omega_{\delta\tau}}
\delta(\vec r_{21}(0)-\vec r) e^{-S},
\end{eqnarray}
where $\Omega_{\tau} = \lambda_{\tau}^{3}$ and $Z$ and $Z_0$ are 
the interacting and non-interacting partition function defined in the previous
Section.
The denominator has been expanded and  written in terms of connected and
disconnected diagrams as in Eq.~\ref{cummulant1}.

\begin{figure} 
    \vskip 0.3 in \begin{center}  
        \subfigure[]{
           \includegraphics[scale=0.35]{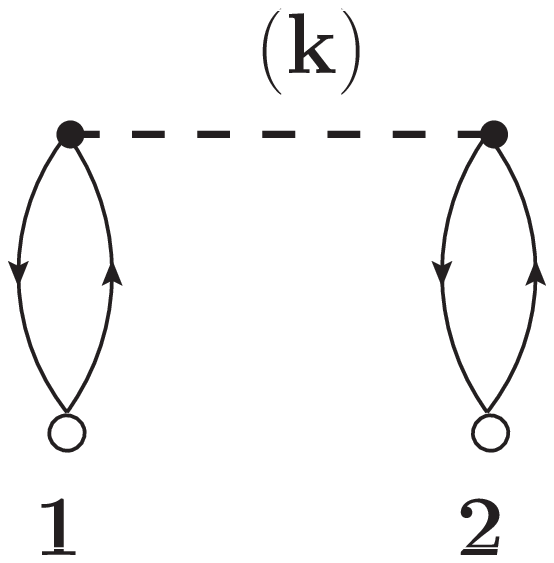}
           \label{fig:ext1}
        }
\hskip 0.2 in
        \subfigure[]{
            \includegraphics[scale=0.35]{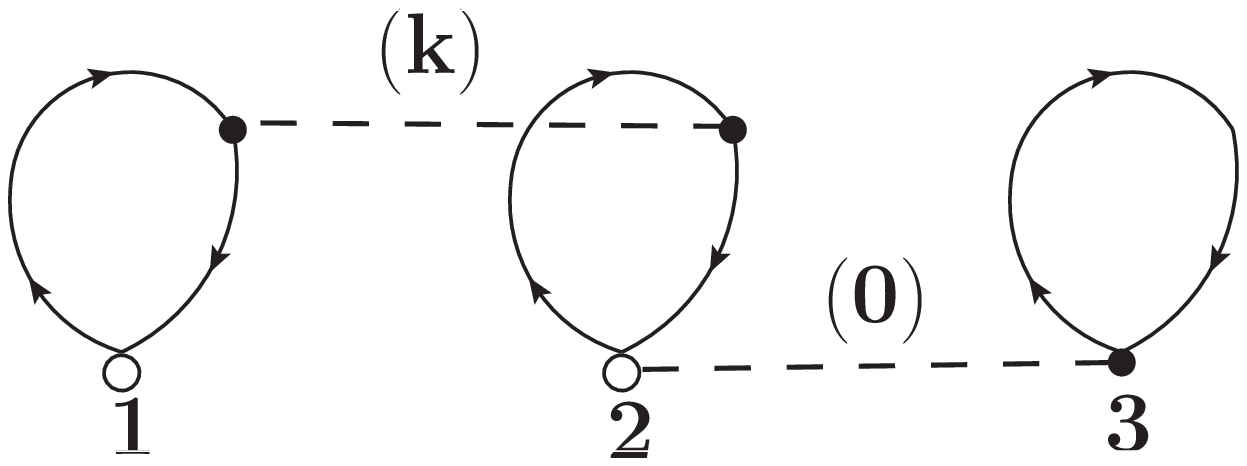}
            \label{fig:ext2}
        }

\vskip 0.2 in
        \subfigure[]{
            \includegraphics[scale=0.35]{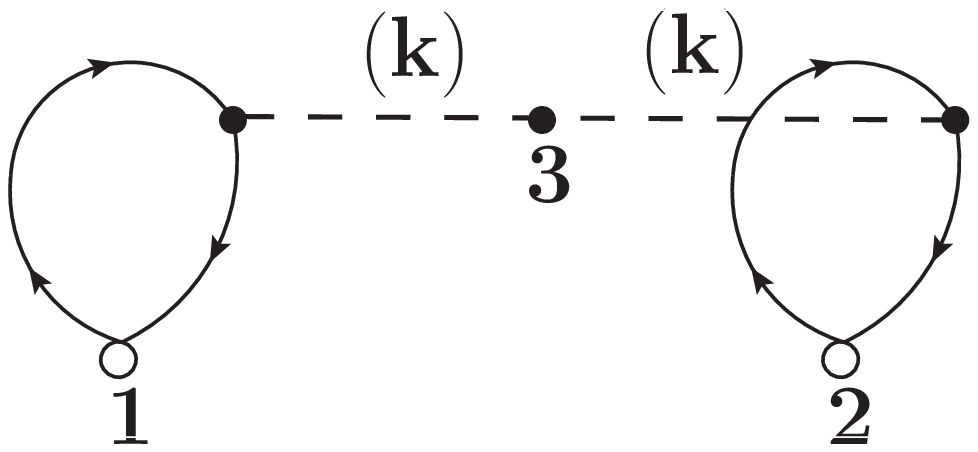}
            \label{fig:ext3}
        }

\caption{\label{fig:numerator} Examples of diagrams contributing 
to the numerator of $g(r)$.}
\end{center}
\label{numerator-diagrams}
\vskip 0.2 in \end{figure}

In order to carry out the cluster expansion of the numerator we need to 
enrich our diagrammatic notation for the numerator diagrams. 
Examples of numerator diagrams are shown in Fig.~\ref{fig:numerator}
The two open circles labeled as $\vec r_1^{(0)}$ and $\vec r_2^{(0)}$ represent
the external points needed in the expression of the numerator.
There are no integrations over these external points
and no $\rho$ factors for their world-lines.  
All other diagrammatic elements and rules are identical to 
those defined in the previous Section. 

Following Wiringa and Pandharipande\cite{RevModPhys.51.821} 
let $[I]$ denote the sum over all connected numerator diagrams $[I]$, i.e., 
diagrams which include points  $\vec r_1^{(0)}$ and $\vec r_2^{(0)}$ as 
external points. When we carry out the cluster expansion of the
numerator we encounter disconnected diagrams with one or more pieces 
and the total expansion of the numerator can be written as
\begin{eqnarray}
{\cal N} = [I] + [I][i] + \frac{1}{2} [I][i][j]  + ...,
\label{cummulantD}
\end{eqnarray}
where a summation over $I$,$i$,$j$, ... is implied. In addition, the
notation $[I][i][j]$ means a disconnected diagram made out of three parts
where no-common particle exists.

Now, we expand the ratio of ${\cal N}/{\cal D}$ and we obtain
\begin{eqnarray}
{{\cal N} \over {\cal D}} &=& [I] + [I][i]-{[I]}\times[i] + {1 \over 2}
[I][i][j] - {1 \over 2} [I]\times[i][j] \nonumber \\
&+& [I]\times[i]\times [j] -[I][i]\times[j] + ... 
\label{expansion}
\end{eqnarray}
The following equations yield the product of sums of diagrams which is contained
in right-hand-side of the above equation:
\begin{eqnarray}
\left [I \right ]\times [i] &=& 
[I][i] + [{\overset{\topleft} {I][i}}] +  ...,  \\
\left [I \right ]\times [i][j] &=& [I][i][j] + 2 [{\overset{\topleft} {I][i}}][j] + 
[\overset{\largetopbracket}{\overset{\topleft} {I][i} ][j]}  \nonumber \\
&+& 
2 [{\overset{\topleft} {\overset{\topleft} {I][j}}}][i] + ...,
\label{overhead1} \\
  \left [I\right ]\times [i]\times [j] &=& [I][i][j] + 2 [{\overset{\topleft} {I][i}}][j] + 
[I][{\overset{\topleft} {j][i}}]  \nonumber \\
&+& [\overset{\largetopbracket}{\overset{\topleft} {I][i} ][j]} +
2 [\overset{\topleft} {I][} i \overset{\topleft} {][j}] +
2 [{\overset{\topleft} {\overset{\topleft} {I][i}}][j}] \nonumber \\
&+& [I][{\overset{\topleft} {\overset{\topleft} {i][j}}}] + 
\overset{\topleft\topright} {[I][i][j]} + ...,
\label{overhead2} \\
\left [I\right ][i]\times [j] &=& [I][i][j] + 
[\overset{\largetopbracket}{I][i][j]} +
[I][\overset{\topleft} {i][j}] +
[\overset{\largetopbracket}{I][\overset{\topleft} {i][j}]}  \nonumber \\
&+& 
[\overset{\largetopbracket} {\overset{\largetopbracket} {I][i][j}}] +
[I][\overset{\topleft} {\overset{\topleft} {i][j}}] + ...,
\label{overhead3}
\end{eqnarray}
where $\overset{\topleft\topright} {[I][i][j]}$, means that the three
pieces share the same particle. The ellipses stand for more than three
disconnected pieces. In addition, the ellipses stand for terms 
in which the number of common particles
is equal to or more than the number of disconnected pieces. 
These latter terms have been neglected because their contribution relative
to that of $I$ vanish in the $N \to \infty$ limit.
 Using the above equations we can write the expression given 
by Eq.~\ref{expansion} as follows:
\begin{eqnarray}
{{\cal N} \over {\cal D}} &=& [I] - [{\overset{\topleft} {I][i}}] 
+{1\over 2}{[\overset{\topleft}{\overset{\topleft}{I][i}][j}]} 
+[\overset{\topleft}{I][} i \overset{\topleft}{][j}]\nonumber \\
&+& \overset{\topleft\topright} {[I][i][j]} + ... .
\label{expansion2}
\end{eqnarray}
We have neglected terms which have the same or more number of overhead lines as the number of disconnected pieces.
Again, as long as each term has just one overhead line less than the number of its disconnected pieces, the contribution 
of the term is of the same order as the order of $[I]$, which is of the order of unity in the case of the distribution function. 

\begin{figure}
    \vskip 0.3 in \begin{center}
        \subfigure[]{
           \includegraphics[scale=0.35]{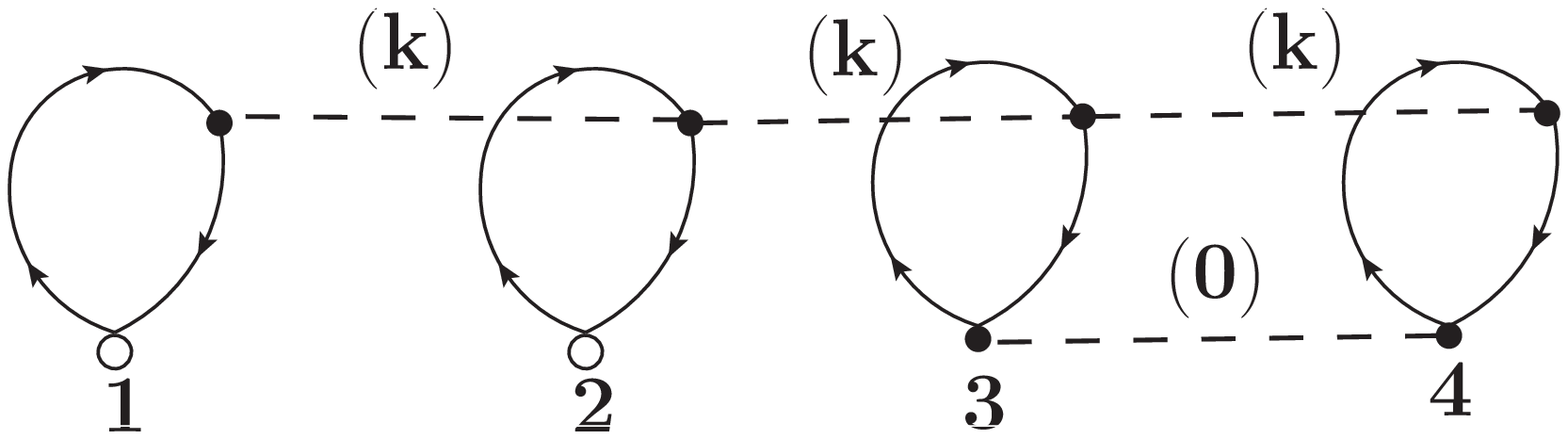}
           \label{fig:factorizable_ext2}
        }
\hskip 0.2 in
        \subfigure[]{
            \includegraphics[scale=0.35]{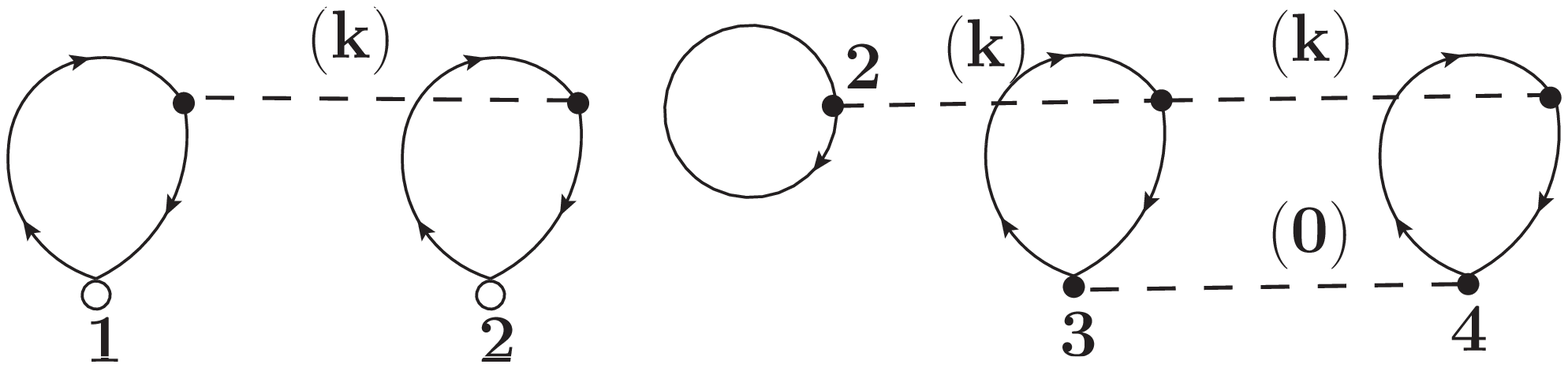}
            \label{fig:factorized_ext2}
        }
\vskip 0.2 in
        \subfigure[]{
            \includegraphics[scale=0.35]{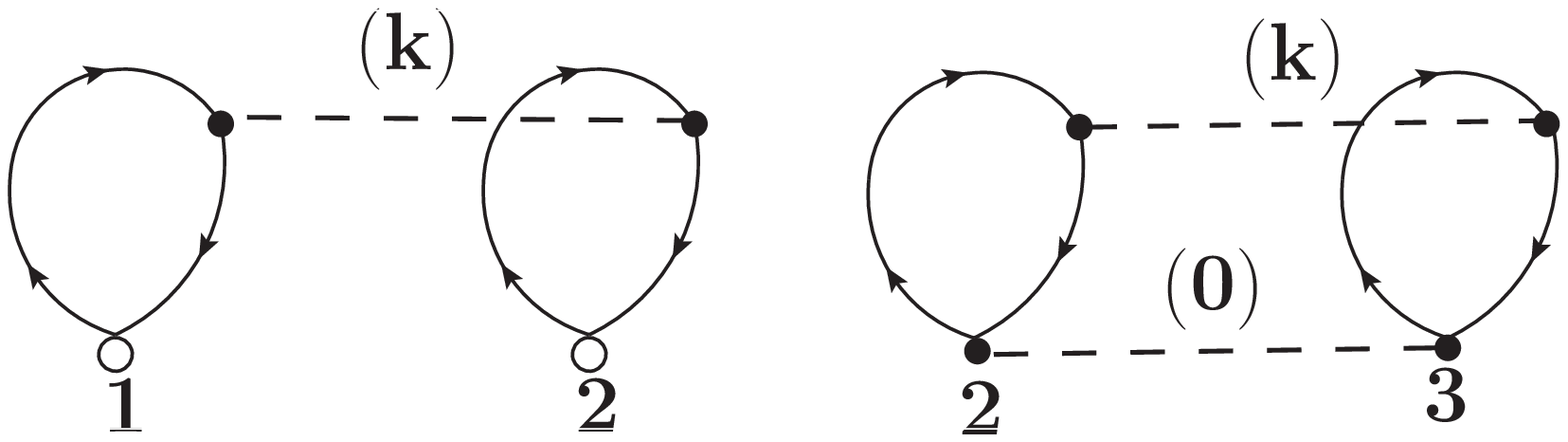}
            \label{fig:factdenom}
        }
\vskip 0.2 in
        \subfigure[]{
            \includegraphics[scale=0.35]{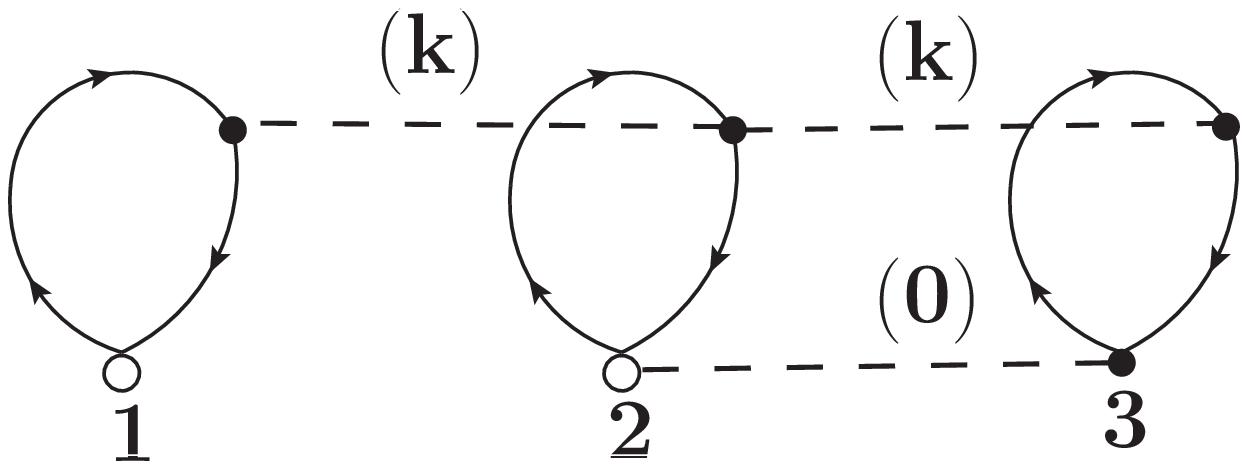}
            \label{fig:nonfactnumerator}
        }
\caption{\label{fig:Aa} Examples of diagrams 
from the families of the
first line in Eq.~\ref{eqn:factorizable-cancellation}.
}
\end{center}
\vskip 0.2 in \end{figure}

We now make use of the factorizability of diagrams to further simplify the above expression. We use the notation $[A]$ to denote a non-factorizable numerator type diagram (this contains two external points) and use $[a]$ to denote a non-factorizable denominator type diagram. We expand our notation to describe this as follows. The following is an example of a connected diagram 
$[I_3]$ which has three factorizable pieces
\begin{eqnarray}
[I_3]=[\overset{\topleft\hskip 0.02 in \topleft}{A \hskip 0.1 in a_1 \hskip 0.1 in a_2}],
\end{eqnarray}
where, the overhead lines denote the points where the diagram $[I_3]$ is factorizable. This means that the sum of the connected diagrams $[I]$ and $[i]$ can be written as follows:
\begin{eqnarray}
[I]&=&[A] + [\overset{\topleft}{A \hskip 0.1 in a}] + [\overset{\topleft\hskip 0.02 in \topleft}{A \hskip 0.1 in a \hskip 0.1 in b}] + {1\over 2}[\overset{\largetopbracket}{\overset{\topleft} {A\hskip 0.1 in a\hskip 0.1 in }b]}\nonumber \\
&+&{1\over 2}{\overset{\threeleft\threeright}{[A\hskip 0.1 in  a\hskip 0.1 in  b]}} + ... , \\
\left [i \right ]&=& [a] +
 {1\over 2}[{\overset{\topleft}{a\hskip 0.1 in b}}] + ... .
\end{eqnarray}
With this, the final expression for the pair distribution function becomes:
\begin{eqnarray}
{\cal{N}\over\cal{D}}&=&[A] + \Bigl ([\overset{\topleft}{A\hskip 0.1 ina}]-[\overset{\topleft}{A][a}]\Bigr )\nonumber \\
&+&\Bigl ([\overset{\topleft\hskip 0.02 in \topleft}{A \hskip 0.1 in a \hskip 0.1 in b}]-[\overset{\topleft}{A\hskip 0.1 in a}\overset{\topleft}{][\hskip 0.01 in b}]-[\overset{\topleft}{A][a}\overset{\topleft}{b}]+[\overset{\topleft}{A][a}
  \overset{\topleft}{][b}]\Bigr )\nonumber \\
&+&\Bigl ({1\over 2}[\overset{\largetopbracket}{\overset{\topleft} {A\hskip 0.1 in a}{\;}\hskip 0.02 in b]}-[\overset{\largetopbracket}{\overset{\topleft} {A\hskip 0.1 in a}][b]}+ 
{1\over 2}[\overset{\largetopbracket}{\overset{\topleft} {A][a}][b]} \Bigr )\nonumber \\
&+&\Bigl ({1\over 2}{\overset{\topleft\topleft}{[A\hskip 0.1 in  a  \hskip 0.1 in b]}} - {\overset{\threeleft\threeright}{[A\hskip 0.05 in  a][b]}}-
{1\over 2}{\overset{\threeleft\threeright}{[A][a\hskip 0.05 in  b]}}+{1\over 2}{\overset{\threeleft\threeright}{[A][a][b]}}\Bigr )\nonumber \\
&+&  ... .
\label{eqn:factorizable-cancellation}
\end{eqnarray}
The terms grouped in each parenthesis cancel exactly in the classical case and we are left with just $[A]$ the sum 
of all the connected and non-factorizable diagrams. In addition, in our 
quantum case, in many cases of diagrams they also cancel.
Fig.~\ref{fig:factorizable_ext2} and Fig.~\ref{fig:factorized_ext2} show typical examples of diagrams from sets 
$[\overset{\topleft}{A\hskip 0.1 ina}]$ and $[\overset{\topleft}{A][a}]$ 
respectively which 
cancel each other. The reason for this cancellation is the following. First, the world-line of particle 2 in the diagram which comes from $[a]$ in this example, is a constant factor, i.e., $L_2(0M)=1/V_{\theta}$. If we erase this ``trivial'' world-line and replace it by
this factor, the two pieces of the diagram of Fig.~\ref{fig:factorized_ext2} become the pieces of the factorizable
diagram of Fig.~\ref{fig:factorizable_ext2} when factorized at point 2.
In fact, any diagram from the $[\overset{\topleft}{A\hskip 0.1 ina}]$ group has a counterpart in the $[\overset{\topleft}{A][a}]$ group
and they mutually cancel. 
The reverse is not true. Namely,
there are diagrams in the $[\overset{\topleft}{A][a}]$ group which do not
have a counterpart in the $[\overset{\topleft}{A\hskip 0.1 in a}]$ group
and, they remain. For example the diagram illustrated in 
Fig.~\ref{fig:factdenom} has no counterpart in the $[\overset{\topleft}{A\hskip 0.1 ina}]$ group,
its counterpart is in the $[A]$ group and it is the
connected-non-factorizable diagram shown in Fig.~\ref{fig:nonfactnumerator}.
In the high temperature limit the diagram in Fig.~\ref{fig:nonfactnumerator}
becomes factorizable (because the world-lines collapse as discussed in the
following Section) and cancels the diagram in Fig.~\ref{fig:factdenom}.
Because pairs of diagrams of this type, such as 
the two diagrams of Fig.~\ref{fig:nonfactnumerator} and 
Fig.~\ref{fig:factdenom}, nearly cancel at even intermediate temperature
we need to either include their contribution together or neglect both.

In general consider any product of diagrams from the $[\overset{\topleft}{A][a}]$ group.
In order to identify its corresponding ``partner'' diagram
in the $[\overset{\topleft}{A\hskip 0.1 ina}]$ group
we join these disconnected pieces $[A]$ and $[a]$ together at the 
common particle to create its corresponding factorizable diagram.
If the created node is a node of  two ``non-trivial'' world-lines
the latter diagram does not exist  in  
the $[\overset{\topleft}{A\hskip 0.1 ina}]$ group, i.e., as a factorizable diagram at the same point. 
Again, by a ``trivial'' world-line we mean those in which 
the corresponding particle has no interactions at any other time-slice
and, therefore, they have been integrated out, yielding a constant factor of $1/V_{\theta}$.

\begin{figure}
    \vskip 0.3 in \begin{center}
        \subfigure[]{
           \includegraphics[scale=0.35]{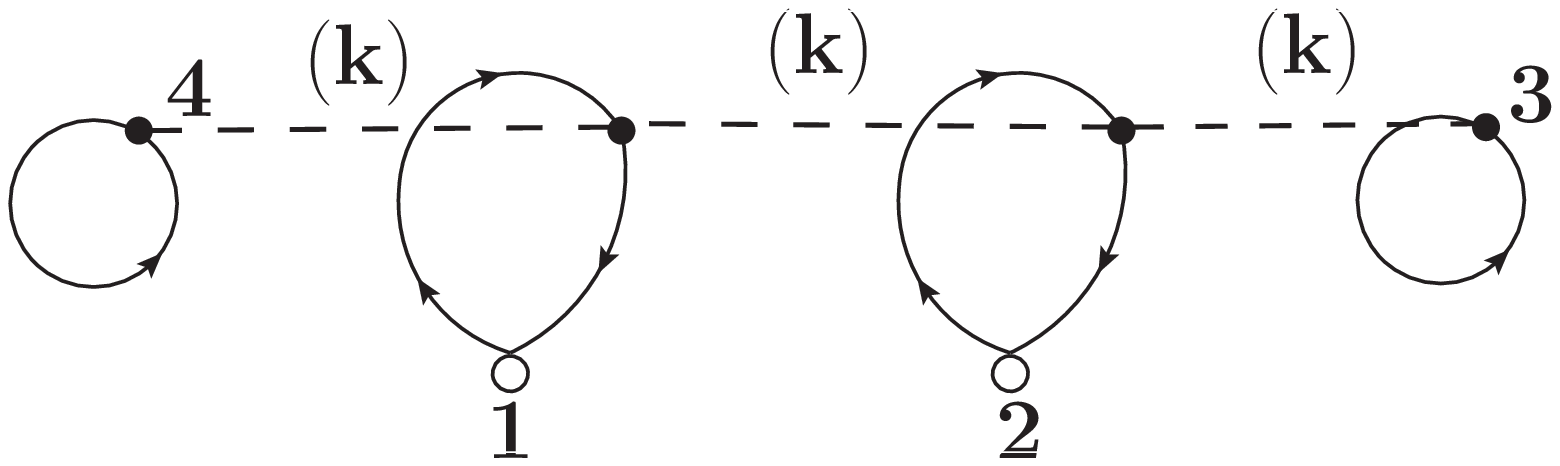}
           \label{fig:a-A-b}
        }
\hskip 0.2 in
        \subfigure[]{
            \includegraphics[scale=0.35]{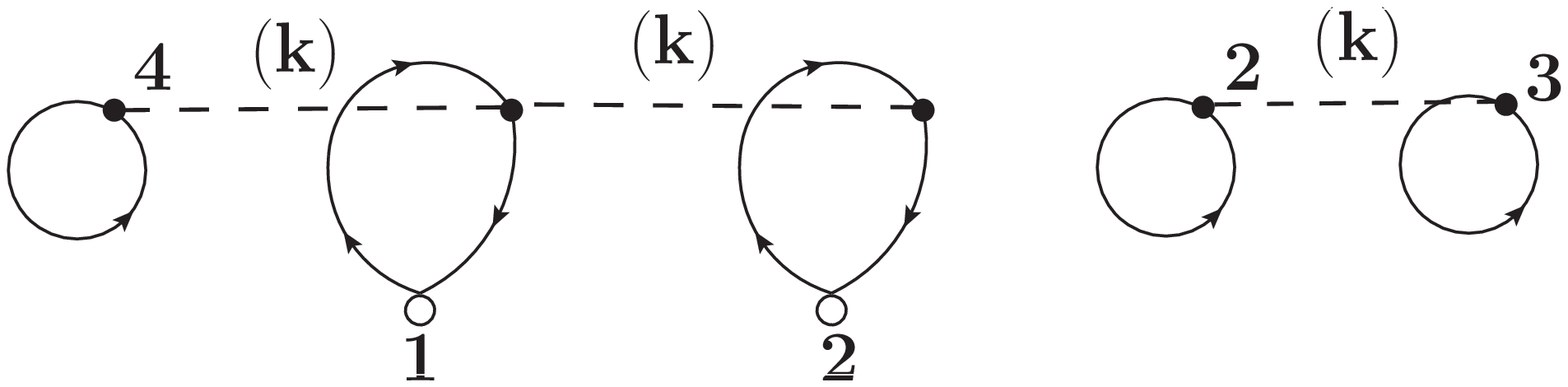}
            \label{fig:a-A-b_counter1}
        }
\vskip 0.2 in
        \subfigure[]{
            \includegraphics[scale=0.35]{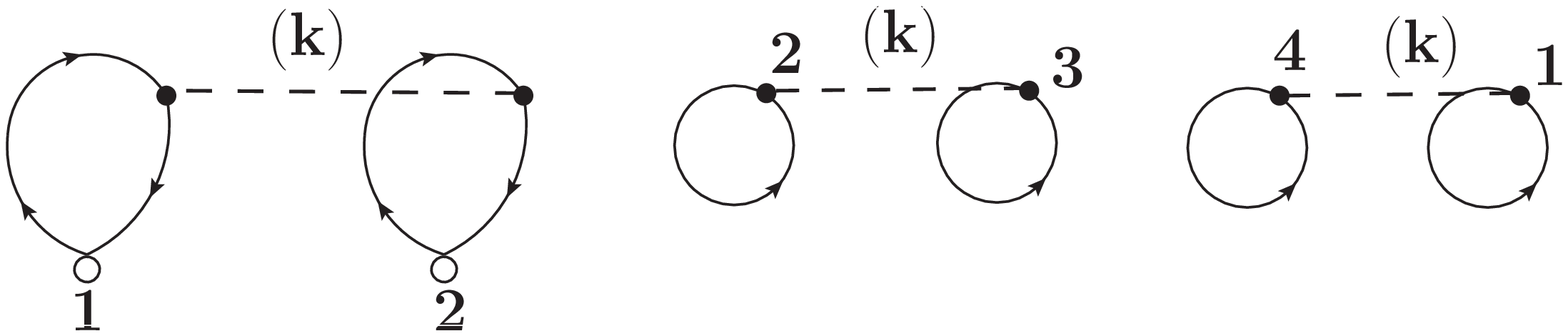}
            \label{fig:a-A-b_counter2}
        }

\caption{\label{fig:aAb} Example of diagrams from the families of the
third line in Eq.~\ref{eqn:factorizable-cancellation} with three factorizable parts which cancel.}
\end{center}
\vskip 0.2 in \end{figure}

In Fig.~\ref{fig:aAb} we give an example of diagrams from the third line
in Eq.~\ref{eqn:factorizable-cancellation} which cancel each other out.
The diagram in Fig.~\ref{fig:a-A-b} is an example from the 
$[\overset{\largetopbracket}{\overset{\topleft} {A\hskip 0.1 in a}{\;}\hskip 0.02 in b]}$ group whereas the diagram in Fig.~\ref{fig:a-A-b_counter1} and 
in Fig.~\ref{fig:a-A-b_counter2} give its two counterparts which correspond to 
the families $[\overset{\largetopbracket}{\overset{\topleft} {A\hskip 0.1 in a}][b]}$ and $[\overset{\largetopbracket}{\overset{\topleft} {A][a}][b]}$. These
three diagrams together cancel out when we take their prefactors and their symmetry factors into account.

\section{Summation methods}
\label{summation}

As discussed in the previous section all disconnected diagrams contributing to $g(r)$, which are products of $[A]$ with $[a]$, $[b]$, ...,
with common particles, have counterparts in either $[A]$ class or in the class of factorizable diagrams.
In addition, we discussed how we can define the partner
of any such disconnected diagram. These ``paired'' diagrams either cancel exactly or they do so in the high temperature
limit. For simplicity we neglected the contribution of all such ``paired'' diagrams because their combined contribution
is very small at moderate temperatures due to this cancellation.

In order to find out the degree of accuracy of the method we choose to test it using particles interacting with the Lennard-Jones
potential
\begin{eqnarray}
v(r)=4\epsilon \Bigl ((\frac{\sigma}{r})^{12}-(\frac{\sigma}{r})^{6}\Bigr ),
\end{eqnarray}
as applied to $^4$He and for particles having the $^4$He atomic mass.
For the case of distinguishable particles we can obtain exact results for the
pair distribution function using the path-integral Monte Carlo method.

\subsection{Density expansion and effective potential}
\label{sec:low-density}
One of the simplest approach would be to expand $g(r)$ and include all the diagrams up to a certain order in $\rho$. 

\begin{figure}[htp]
\vskip 0.3 in
\begin{center} 
\includegraphics[scale=0.35]{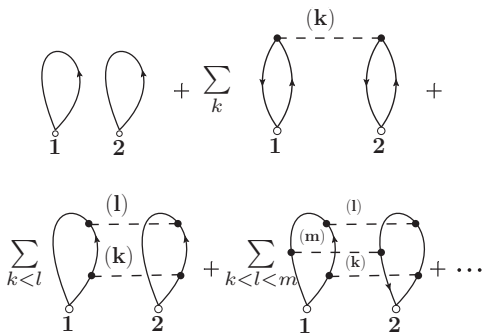}
\caption{\label{fig:zeroth} Expansion of $g(r)$ up to zeroth order 
in density.}
\end{center}
\end{figure}

(a) {\bf Zeroth order}:
The sum of zeroth order diagrams is given by the infinite series
shown in Fig.~\ref{fig:zeroth}. It is the sum of all possible
two-body diagrams which can be
obtained by considering all possible ways in which $h$-lines connect
the coordinates of these two particles at any time-slice.

\begin{figure}[htp]
\vskip 0.3 in
\begin{center} 
\includegraphics[scale=0.30]{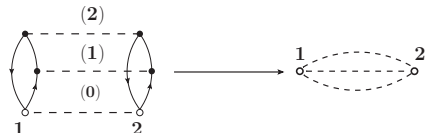} 
\caption{Collapse of quantum world-line in the high temperature limit\label{fig:collapse}. The diagram on the left is also an example of a ladder diagram. }
\end{center}
\end{figure}

The sum of these diagrams is needed for two important reasons.
We need to sum the entire series in order to obtain the
correct behavior in low density limit.
Second we need the entire series in order to obtain the
correct classical limit at high temperature.
To understand the latter, consider for simplicity the case of
fixed number of time-slices and the expression for $Z$ given 
by Eq.~\ref{part1}. When the temperature is high,  
the world-line of any particle collapses to a point: 
this can be realized mathematically by noticing that the world-line is made up of the product of the Gaussians of the form given by Eq.~\ref{world-line} 
and as the temperature becomes high,
$\delta \tau = \hbar\beta/M \to 0$, these Gaussians approach
a delta-function of the difference in the two positions of the particle
at two successive imaginary-time slices. In this case, the
integrations over  $\vec r^{(k)}_i$ for all $k\ne 0$ in 
Eq.~\ref{part1} can be carried out. This eliminates all the integrals
corresponding to the coordinates $\vec r^{(k)}_i$ for $k\ne 0$, and sets $\vec r^{(k)}_i=\vec r^{(0)}_i$ for $k\ne 0$ in the  
integrand, thus, $Z$ contains just  $N$ integrals
over the $N$-particle coordinates $\vec r^{(0)}_i$:
\begin{eqnarray}
{ Z \over {Z_0}} &=& \int 
\prod_{n=1}^N\frac{d^{3}r_n^{(0)}}{V} \prod_{i<j}(1+h_{ij}^{(0)})^M.
\label{part-high-temp}
\end{eqnarray} 
Notice that, using the definition of $h^{(0)}_{ij}$ (Eq.~\ref{h-factor}) the above
factor becomes
\begin{eqnarray}
(1+h_{ij}^{(0)})^M = e^{-\beta v(r^{(0)}_{ij})},
\label{classical-limit-1}
\end{eqnarray}
i.e., we recover the classical partition function. However, since this product has been expanded to obtain the cluster expansion in the quantum case, i.e., 
\begin{eqnarray}
 e^{-\beta v(r^{(0)}_{ij})} = (1+h^{(0)}_{ij})^M = \sum_{k} \Bigl (\begin{array}{c} M   \\
    k \end{array} \Bigr ) (h^{(0)}_{ij})^k,
\label{classical-limit-2}
\end{eqnarray}
it implies that diagrammatically the collapse of the world-line leads
to multiple $h$-lines connecting any two particles at 
the coordinates which correspond to the initial time.
 Fig.~\ref{fig:collapse} shows this diagrammatically for the case of $M=3$.

\begin{figure}[htp]
\vskip 0.3 in
\begin{center} 
\includegraphics[scale=0.35]{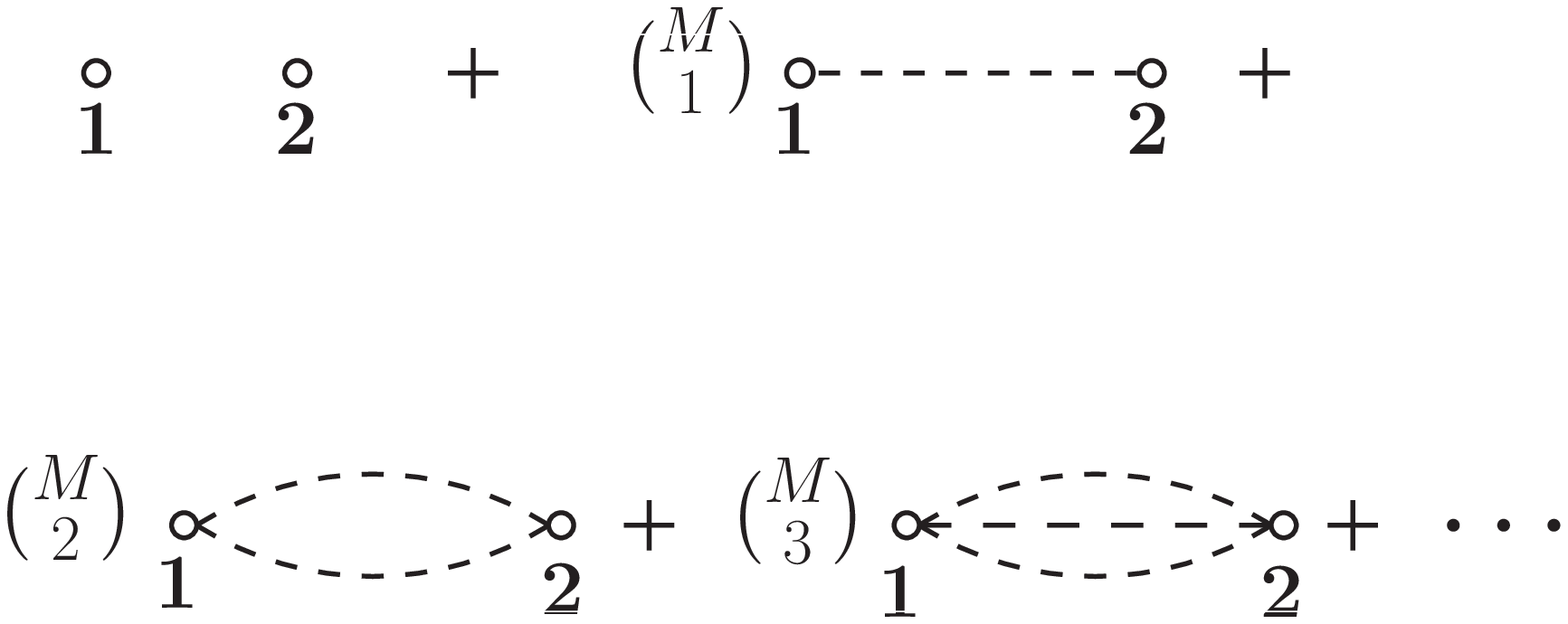}
\caption{The fate of the series
of the zeroth-order diagrams of Fig.~\ref{fig:zeroth} in the high-temperature
limit.
\label{fig:demonstration}}
\end{center}
\end{figure}

In Fig.~\ref{fig:demonstration} we show the fate of the series
of the zeroth-order diagrams of Fig.~\ref{fig:zeroth} in the high-temperature
limit.

This explains the importance of including all zeroth-order diagrams 
in our formalism. The sum $g_0(r_{12})$ of the zeroth-order diagrams contributing
to $g(r)$, i.e., those in Fig.~\ref{fig:zeroth}, defines an effective 
potential $v_e(r)$ as follows:
\begin{eqnarray}
  g_0(r_{12}) = \exp \bigl (-\beta v_{e}(r_{12}) \bigr ).
  \label{eq:effective-potential}
\end{eqnarray}
In order to justify this definition, first, notice that the
bare potential is obtained from the 
high-temperature  (classical) 
and zero-density limit of the distribution function, which is given as: 
\begin{eqnarray}
  \lim_{\beta \to 0}  g_0(r_{12}) = \exp \bigl (-\beta v(r_{12}) \bigr ).
  \label{eq:classical-distribution}
\end{eqnarray}
Also, the high temperature limit of the sum presented in Fig.~\ref{fig:zeroth}
is the sum given in Fig.~\ref{fig:demonstration}. The latter sum is equal to
the result given by the above Eq.~\ref{eq:classical-distribution}
This implies that in the above definition the effective potential
corresponds to the
case where instead of freezing the particles coordinates
at their initial values, which would lead to the classical limit,
we  allow them to fluctuate in imaginary-time. In the zero-density limit,
there are only the world-lines of the two external particles that
matter.

The calculation  of the sum of  all zeroth-order diagrams can  be done
easily by noting that the diagonal  part of the exact two body density
matrix is  directly proportional  to the sum  of all  the zeroth-order
diagrams in  the density  expansion of $g(r)$.  This is true provided
that sufficiently large  number of time slices have been  used to find
the sum. This sum  can be calculated by using the matrix
squaring method\cite{Elba,Aust.J.Phys.26}  
for the two-body  density matrix. The exact  two-body
density matrix at  any temperature can be calculated  by starting from
the exact two-body density matrix  at a very high temperature and then
using the matrix squaring  method to obtain the
exact density matrix at lower temperature.

\begin{figure}
    \begin{center}
        \subfigure[]{
           \includegraphics[scale=0.32]{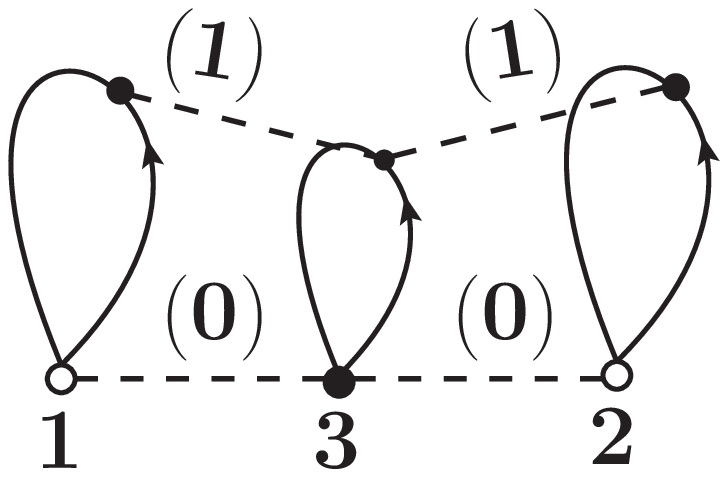}
           \label{fig:first1}
        }
\hskip 0.6 in
        \subfigure[]{
            \includegraphics[scale=0.32]{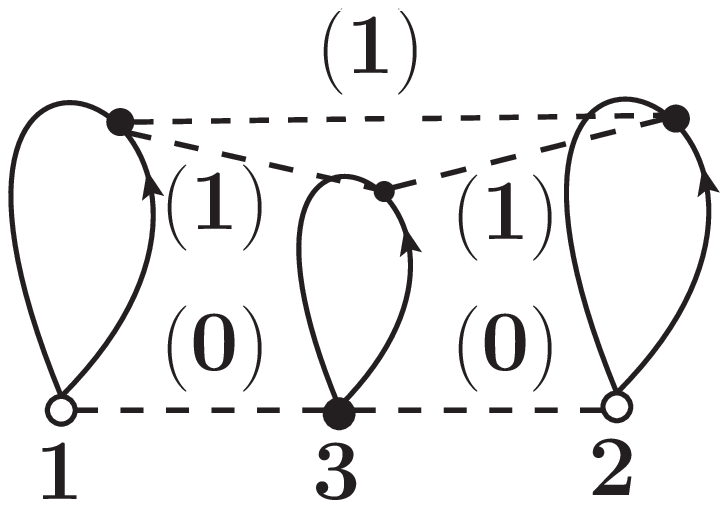}
            \label{fig:first2}
        }
    \end{center}
    \caption{Fig.~(a) 
      and Fig.~(b) are examples of $1^{st}$ (in $\rho$) order ladder diagrams.}
\label{fig:first-order}
\end{figure}

\begin{figure}[H]
\vskip 0.2 in
    \begin{center}
        \subfigure[]{
           \includegraphics[scale=0.32]{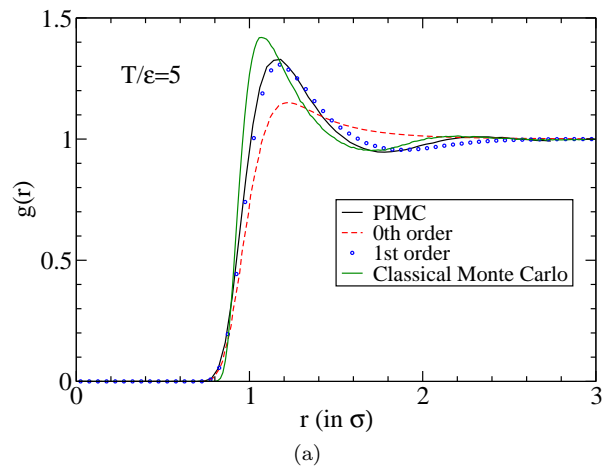}
           \label{fig:sch1_5t}
        }
\vskip 0.3 in
        \subfigure[]{
            \includegraphics[scale=0.45]{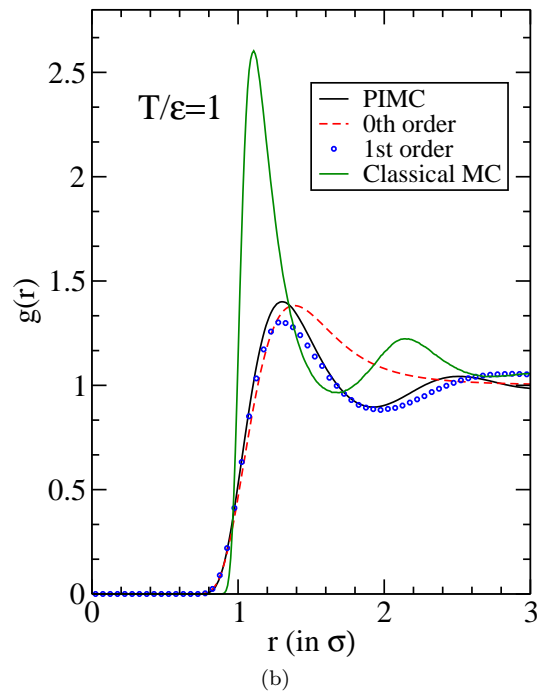}
            \label{fig:sch1_1.5t}
        }

    \end{center}
    \caption{Comparison of the calculated $g(r)$ within zeroth order (red dashed-line) and first order (blue circles) 
with the results of our PIMC simulation (black-line) for $T/\epsilon=5$ (panel (a)) and $T/\epsilon=1$ (panel (b)). 
The green solid-line is the result of
      classical Monte Carlo simulation. }
\end{figure}

(b) {\bf First Order:} Unlike the zeroth-order diagrams 
the higher order ladder diagrams 
for an arbitrary number of time slices are harder to calculate.
Examples of diagrams which are first order in $\rho$
are shown in Fig.~\ref{fig:first-order}.
We expect that such an expansion up to first order in $\rho$ 
will only give accurate results in the low density regime and also in the
high temperature regime.
To compare the first order correction against the PIMC we have used only two time 
slices but using the $v_{eff}(r)$ instead of the bare interaction calculated at $\beta/2$.
Using this approach we calculated $g(r)$ at both low temperature ($T=1$) and at 
high temperature ($T=5$) and for helium density $\rho=0.365 \sigma^{-3}$. 
As we discuss in Appendix~\ref{sec:convergence}, while the effective interaction
provides a good approximation for low-density for $M=1$, when the value of $M$
is increased, the effective interaction approach yields results 
which at first diverge from the exact solution at low $\rho$.
On the contrary, for helium density $\rho=0.365 \sigma^{-3}$,  the $g(r)$ obtained for $M=2$ using the effective interaction in the PIMC simulation 
is close to that corresponding to the
$M \to \infty$ limit.  In the Fig.~\ref{fig:sch1_5t} and Fig.~\ref{fig:sch1_1.5t} we compare the calculated $g(r)$ 
with the results of PIMC. This approximation surprisingly yields results 
for $g(r)$, which, in general, agree with the PIMC results at smaller distances at both low and high 
temperature. As expected, the agreement with Monte Carlo is better when the temperature is high.

\subsection{Quantum Hypernetted Chain Equations (QHNC)}
\label{sec:QHNC}
Next, we will attempt to find the equations which will allow us to resum 
the diagrams contributing to $g(r)$ by generalizing the 
classical HNC approach to
the quantum case. In order to achieve this, we begin by 
first classifying the diagrams as nodal ($N$) diagrams or as non-nodal 
(or composite) ($X$) diagrams (this is similar to
the classical case\cite{hiroike1,hiroike2,morita1,morita2,morita3,morita4,morita5}) 
and we ignore the elementary diagrams. 
We make use of the fact that any nodal diagram is either 
a convolution of a non-nodal diagram with another non-nodal diagram 
or a convolution of a non-nodal diagram with a nodal diagram. 
To be able to sum all the diagrams which contribute to $g(r)$ we will need
to consider convolutions of auxiliary diagrammatic pieces which do not
necessarily contribute to $g(r)$ by themselves directly, but only through convolution with other diagrams or sub-diagrams.
Such sub-diagrams can have external points which 
can be extremities of only $h$ lines or there can also be sub-diagrams
in which one or both of their external points lie on the path 
of an $L$-line. Because of this, we further divide the
class of the $N$ and $X$ diagrams into sub-groups, $N_{\alpha\beta}$
and $X_{\alpha\beta}$ where $\alpha$ and $\beta$ take the values
$h$ or $w$ depending on whether the external point is an extremity
of just $h$ lines or the external point lies on the path of an $L$-line.  
This is formally somewhat similar to the case of the so-called 
Fermi HNC (FHNC) \cite{Fantoni1975,PhysRevB.28.3770} equations technique,
which was applied
to calculate ground-state expectation values with Jastrow-Slater variational 
wavefunctions. 
In our case, however, unlike the case of FHNC, we need to further divide 
the nodal and composite diagrams into sub-groups of equal-time and unequal-time diagrams. Next, we outline the
definitions of various sub-groups in detail as follows:

1. $N_{hh}$ or $X_{hh}$ diagrams: 
In these type of diagrams the two external points $\vec{r}_1^{(k)}$ and $\vec{r}_2^{(l)}$ are connected by $h$-lines.  When $k$=$l$, these diagrams are at 
equal time and are denoted by $N_{hh}^{(k)}(r_{12})$ and $X_{hh}^{(k)}(r_{12})$. 
When $k \ne l$, these are unequal-time diagrams and are 
denoted by $N_{hh}^{(kl)}(r_{12})$ (or $X_{hh}^{(kl)}(r_{12})$). 
Some examples are shown in Fig.~\ref{fig:hh}:

\begin{figure}
    \begin{center}
        \subfigure[]{
           \includegraphics[scale=0.32]{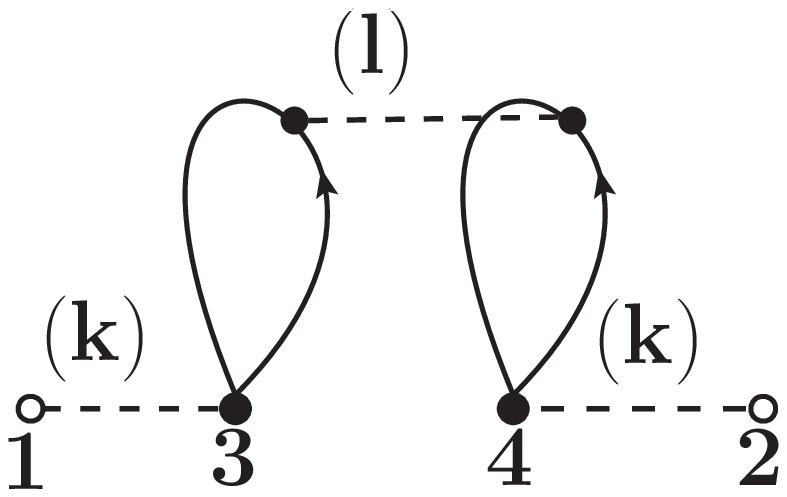}
           \label{fig:nhh(eq)}
        }
\hskip 0.1 in
        \subfigure[]{
            \includegraphics[scale=0.32]{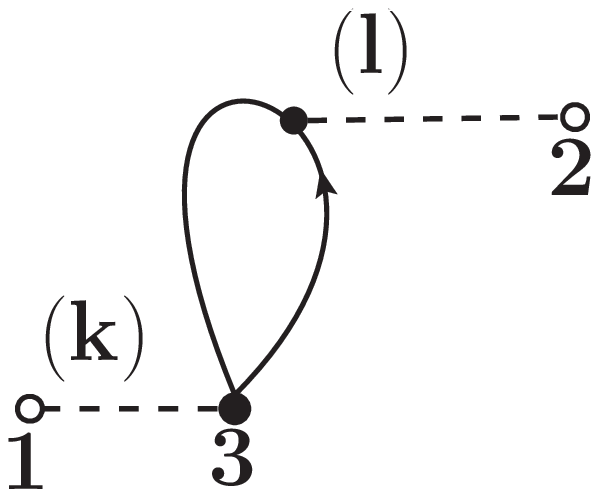}
            \label{fig:nhh(uneq)}
        }
\vskip 0.3 in
        \subfigure[]{
           \includegraphics[scale=0.32]{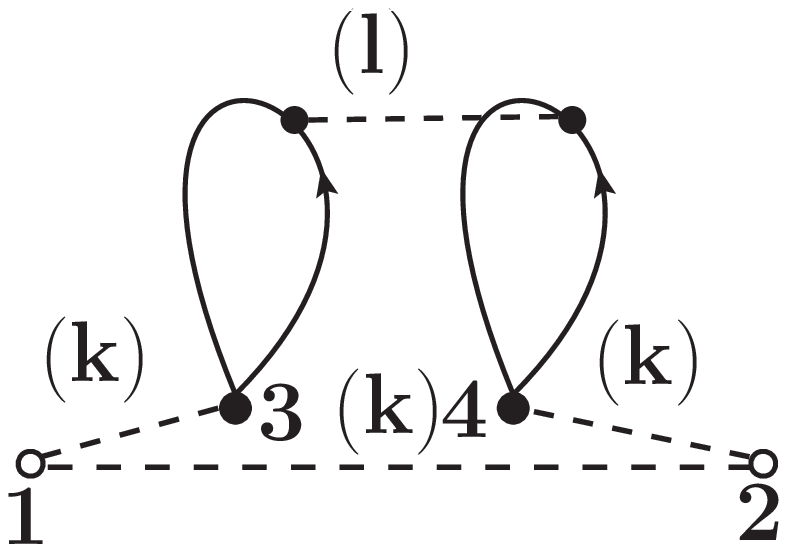}
           \label{fig:xhh(eq)}
        }
\hskip 0.1 in
        \subfigure[]{
            \includegraphics[scale=0.32]{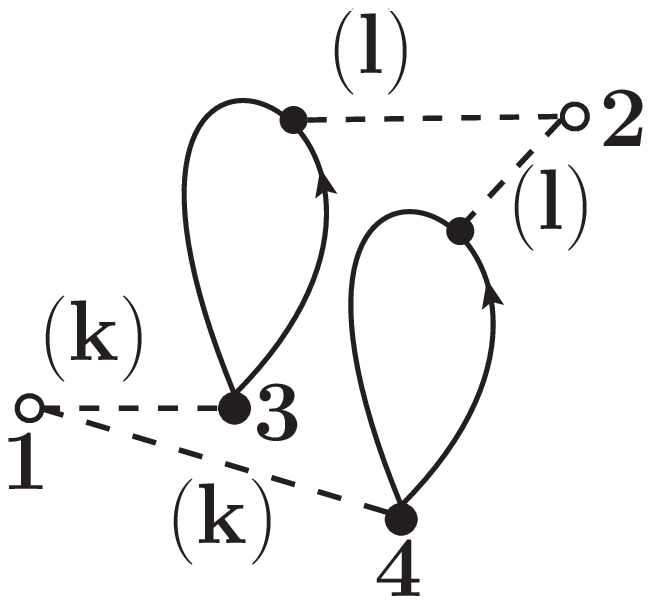}
            \label{fig:xhh(uneq)}
        }

    \end{center}
    \caption{(a) An example of equal-time $N_{hh}$ diagram. (b) An example of a unequal-time $N_{hh}$ diagram.
      (c) An example of an equal-time $X_{hh}$ diagram. (d) An example of a unequal-time $X_{hh}$ diagram. \label{fig:hh}}
\end{figure}

\begin{figure}
    \begin{center}
        \subfigure[]{
           \includegraphics[scale=0.35]{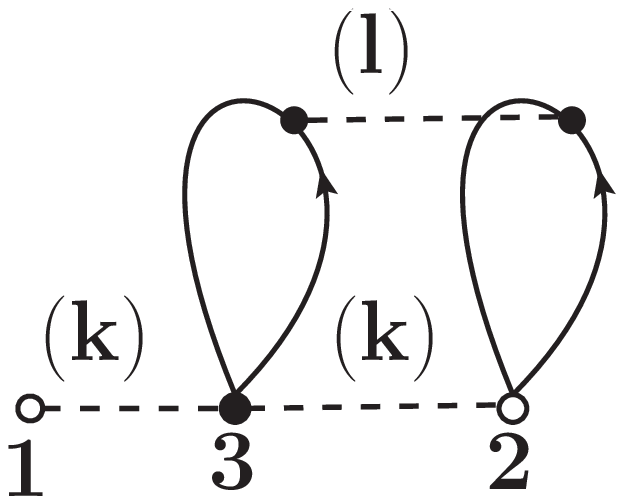}
           \label{fig:nhw(eq)}
        }
\hskip 0.6 in
        \subfigure[]{
            \includegraphics[scale=0.35]{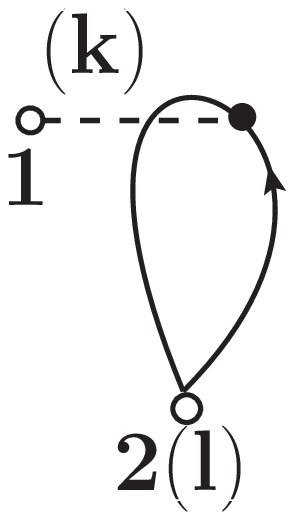}
            \label{fig:nhw(uneq)}
        }
\vskip 0.6 in
        \subfigure[]{
           \includegraphics[scale=0.35]{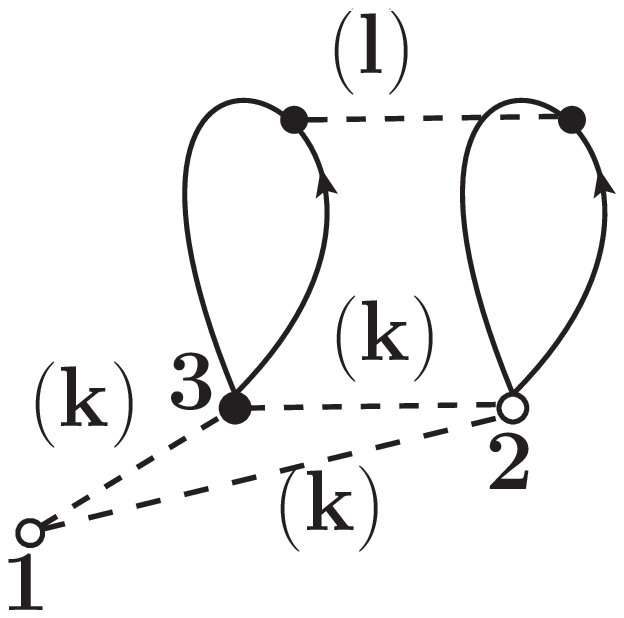}
           \label{fig:xhw(eq)}
        }
\hskip 0.6 in
        \subfigure[]{
            \includegraphics[scale=0.35]{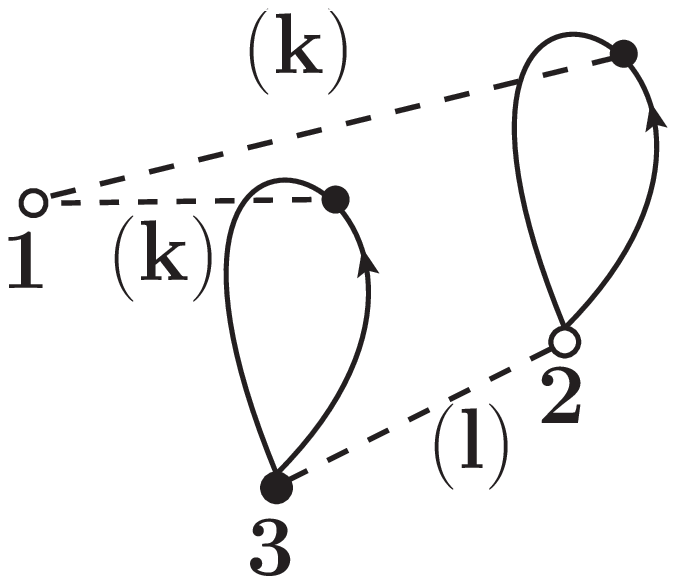}
            \label{fig:xhw(uneq)}
        }

    \end{center}
    \caption{(a) An example of equal-time $N_{hw}$ diagram. (b) An example of a unequal-time $N_{hw}$ diagram.
      (c) An example of an equal-time $X_{hw}$ diagram. (d) An example of a unequal-time $X_{hw}$ diagram. \label{fig:hw}}
\end{figure}

2. $N_{hw}$ or $X_{hw}$ type diagrams: In these type of diagrams the external point $\vec{r}_1^{(k)}$ is connected by only $h$-lines and the 
external point $\vec{r}_2^{(l)}$ lies on a $L$-line.   When $k$=$l$, these are equal time diagrams and
are denoted by $N_{hw}^{(k)}(r_{12})$ (or $X_{hw}^{(k)}(r_{12})$). 
When $k \neq l$,  these are unequal-time diagrams and are denoted by $N_{hw}^{(kl)}(r_{12})$ (or $X_{hw}^{(kl)}(r_{12})$). 
In a similar way we can have  $N_{wh}$ or $X_{wh}$ type of diagrams. Some examples are shown in Fig.~\ref{fig:hw}.

\begin{figure}
    \begin{center}
        \subfigure[]{
           \includegraphics[scale=0.35]{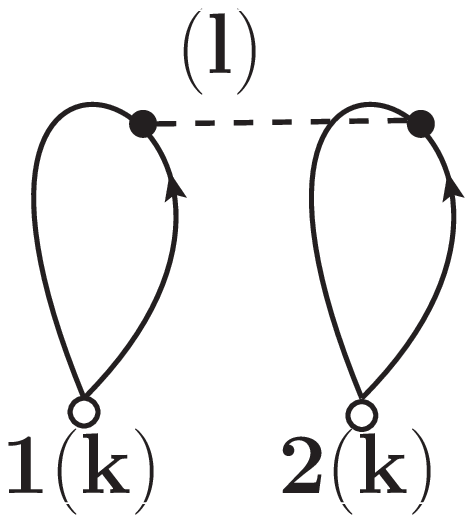}
           \label{fig:nww(eq)}
        }
\hskip 0.6 in
        \subfigure[]{
            \includegraphics[scale=0.35]{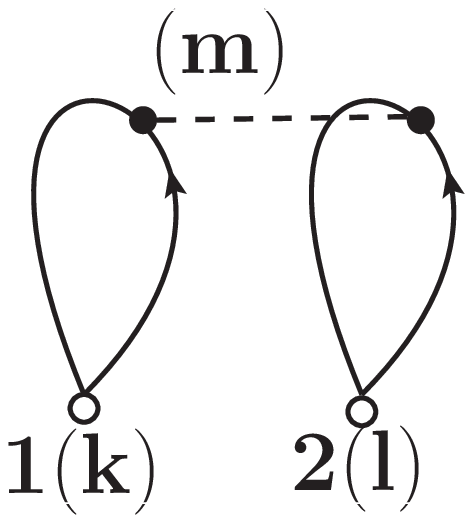}
            \label{fig:nww(uneq)}
        }
\vskip 0.6 in
        \subfigure[]{
           \includegraphics[scale=0.35]{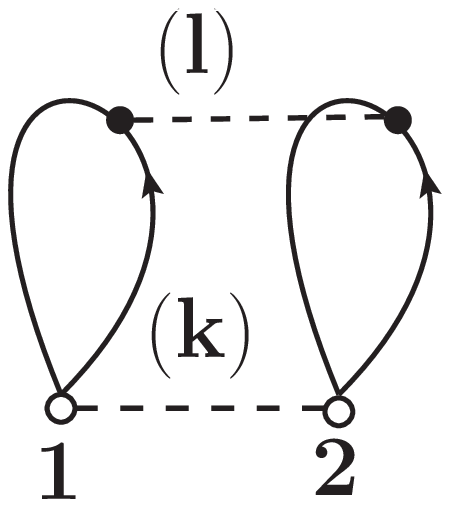}
           \label{fig:xww(eq)}
        }
\hskip 0.6 in
        \subfigure[]{
            \includegraphics[scale=0.35]{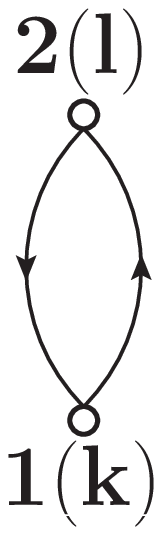}
            \label{fig:xww(uneq)}
        }

    \end{center}
\caption{Fig. (a) shows an example of equal time $N_{ww}$ diagram. Fig. (b) shows unequal-time $N_{ww}$ diagram. Fig. (c) shows an example of equal time $X_{ww}$ diagram. Fig. (d) shows unequal-time $X_{ww}$ diagram.\label{fig:ww}}
\end{figure}

3. $N_{ww}$ or $X_{ww}$ diagrams: 
In these type of diagrams the two external points $\vec{r}_1^{(k)}$ and $\vec{r}_2^{(l)}$ lie on $L$-lines.  When $k$=$l$, these are equal time diagrams 
and are denoted by $N_{ww}^{(k)}(r_{12})$ (or $X_{ww}^{(k)}(r_{12})$). When $k \ne l$ these are unequal-time diagrams and are denoted by $N_{ww}^{(kl)}(r_{12})$
(or $X_{ww}^{(kl)}(r_{12})$). Some examples are shown in Fig.~\ref{fig:ww}.

Next, a set of self-consistent closed equations for various $X$ and $N$ type of diagrams can be written. Any $N_{hh}$ type diagram can be generated as 
a convolution of $hh$ type diagram with another $hh$ type diagram or with an $wh$ type diagram. 
Similarly, a $N_{wh}$ type of diagram may be constructed as a 
convolution of a $wh$ type diagram with another $wh$ type diagram or 
from a $ww$ type of diagram with a $hh$ type diagram. Finally, $N_{ww}$ type of diagram may be obtained as a  convolution of (a) a $ww$ type with a $hw$ type diagram or (b) a $wh$ type with a $hw$ type diagram, or (c) a $wh$ type with a 
$ww$ type diagram. The basic rule involved in writing these equations is 
that any two points can only be convoluted if at least one of them is of $h$-type.
In a similar fashion we can see that an unequal-time diagram can be generated by taking convolution of two unequal-time diagrams or a convolution of an equal time diagram with an unequal-time diagram. 
The generalized equations for any type of nodal diagram are given below:
\begin{eqnarray}
N_{\alpha\beta}^{(kl)}(r_{12})&=&\rho\overset{M-1}{\underset{m=0}{\sum}}\underset{\gamma,\gamma'}{\sum}'\int X_{\alpha\gamma}^{(km)}(r_{13})(X_{\gamma'\beta}^{(ml)}(r_{32})+\nonumber \\
&&N_{\gamma'\beta}^{(ml)}(r_{32}))d^3\vec{r}_3^{(m)}.
\label{eq:qhnc}
\end{eqnarray}
Here, the subscripts $\alpha,\beta,\gamma$ and $\gamma'$ can be either $w$ or $h$ and the presence of $\sum'$ means that the sum over the $\gamma$ and $\gamma'$ indices is constrained such that (as explained before) $\gamma$ and $\gamma'$ both can not be $w$ simultaneously.
The unequal-time $X$-type of diagrams can be written in terms of the unequal-time nodal diagrams in the following way:
\begin{eqnarray}
X_{hh}^{(kl)}(r_{12})=e^{N_{hh}^{(kl)}(r_{12})}-N_{hh}^{(kl)}(r_{12})-1,
\end{eqnarray}

\begin{eqnarray}
X_{hw}^{(kl)}(r_{12})&=&e^{N_{hh}^{(kl)}(r_{12})}N_{hw}^{(kl)}(r_{12})-N_{hw}^{(kl)}(r_{12}),
\end{eqnarray}

\begin{eqnarray}
X_{ww}^{(kl)}(r_{12})&=&e^{N_{hh}^{(kl)}(r_{12})}[N_{ww}^{(kl)}(r_{12})+N_{hw}^{(kl)}(r_{12}) N_{wh}^{(kl)}(r_{12})]
\nonumber \\
&+&\frac{1}{\rho V_{\theta}} l(r_{12},\tau_{kl})\cdot l(r_{12},\theta-\tau_{kl})\nonumber \\
&-&N_{ww}^{(kl)}(r_{12}).
\end{eqnarray}
In the above equations $\tau_{kl}$ is the time interval between 
the $k^{th}$ and $l^{th}$ time slices and the function $l$ is defined as 
follows:
\begin{eqnarray}
l(r_{12},\tau_{kl})&=&\frac{1}{\lambda^3_{\tau_{kl}}}
\exp\Bigl [{-\pi \frac{(\vec{r}_1^{(k)}-\vec{r}_2^{(l)})^2}{\lambda^2_{\tau_{kl}}}}
\Bigr ].
\end {eqnarray}
The equal time $X$-type of diagrams can be written in terms of the equal time nodal diagrams in the following way:
\begin{eqnarray}
X_{hh}^{(l)}(r_{12})&=&f(r_{12})e^{N_{hh}^{(l)}(r_{12})}-N_{hh}^{(l)}(r_{12})-1, \\
X_{hw}^{(l)}(r_{12})&=&f(r_{12})e^{N_{hh}^{(l)}(r_{12})}N_{hw}^{(l)}(r_{12}) 
- N_{hw}^{(l)}(r_{12}), \\
X_{ww}^{(l)}(r_{12})&=&f(r_{12})e^{N_{hh}^{(l)}(r_{12})}[N_{ww}^{(l)}(r_{12})+N_{hw}^{(l)}(r_{12})  \nonumber \\
&\times&N_{wh}^{(l)}(r_{12})]-N_{ww}^{(l)}(r_{12}), \\
f(r_{12}) &=& 1 + h(r_{12}) = e^{-{{\delta \tau} \over {\hbar}} v(r_{12})}.
\end{eqnarray} 
By solving these equations we can find the numerical values of all the nodal diagrams and then use them to find the pair distribution function. The pair distribution can be written in the following way in terms of the nodal functions:
\begin{eqnarray}
g(r_{12})=&&f(r_{12}) e^{N_{hh}^{(0)}(r_{12})}(1+N_{ww}^{(0)}(r_{12})+ \nonumber \\
&&2 N_{hw}^{(0)}(r_{12})+N_{hw}^{(0)}(r_{12}) N_{wh}^{(0)}(r_{12})).
\end{eqnarray}

\subsection{Implementation of QHNC}

\subsubsection{Effective interaction approach}

In our  QHNC equations  we used  the effective  interaction obtained
from  the   exact  two-body  density   matrix  instead  of   the  bare
interaction.  Namely,  instead of  $h^{(k)}_{ij}$ we use  an effective
$h$-line,  i.e., the  $h^{(k)}_{ij}$ obtained  from Eq.~\ref{h-factor}
using $v_e(r)$  (given by  Eq.~\ref{eq:effective-potential}) instead
of  the bare  interaction.

There are two ways to justify the use of the effective interaction
instead of the bare. 

A) Using the exact two-body density-matrix (ETBDM) as starting point: 
We note that we could have defined our original starting point, such
that, instead of using the bare interaction in Eq.~\ref{h-factor}
to use the effective potential. This effective $h$, i.e., $h_e$ and the bare 
$h$ are the same to order $\delta \tau$. 
This is what is done in most PIMC studies\cite{PhysRevB.30.2555,PhysRevLett.56.351,PhysRevB.36.8343,PhysRevB.39.2084,Elba} 
of liquid $^4$He.  Therefore, since we need to make sure that 
the process converges by taking the $M\to \infty$ limit, 
both potentials should yield the same limit. 
The effective potential corresponds to using the ETBDM 
 in the two-body Trotter break-up of the
operator $e^{-\delta\tau\hat H/\hbar }$ which leads
to using our effective potential in Eq.~\ref{h-factor}.
The reason to prefer 
the effective potential as a starting point, is our hope that it
will lead to a faster convergence\cite{PhysRevB.30.2555} with respect to $M$.

B) Using the primitive Trotter approximation (PTA) as 
starting point: This means that we simply begin our cluster expansion or
the PIMC simulation starting from the bare interaction 
in  Eq.~\ref{h-factor}. In this approach
the effective interaction emerges as the zero-density limit or by 
retaining just the two-body clusters.   Furthermore, even if we start from the
PTA as a starting point, this effective interaction should be 
incorporated in the  QHNC equations, because they do not account 
for diagrams    such   as    those   in
Fig.~\ref{fig:zeroth}  which  contain  more  than  two  $h$-lines.  As
discussed  in  Sec.~\ref{sec:low-density}  the inclusion  of  all  the
diagrams  shown in  Fig.~\ref{fig:zeroth} is  necessary to  obtain the
correct classical limit and low density limit. 

The reason that we have the choice between A and B is because
of the fact that $M$ and $\delta \tau$ are tunable.
However, there are different issues with both using the ETBDM or the 
PTA as starting points  to justify the use of the effective interaction.
If we use PTA as starting point, 
the effective interaction is derived as the zero density
limit. Within this approach, if we simply 
replace the $h$-line with an effective $h$-line it can lead
to a consistency problem which is discussed in the next subsections.
If we start from the ETBDM, there is no such conceptual issue with
justifying its use; 
however, as we will see in the next two subsections, there is a different
issue which arises with the {\it convergence with $M$} of the PIMC method 
in the low-density regime.

In the next subsection, we have solved the  QHNC equations, using the
effective potential for the case of $M=1$ and $M=2$, to find the pair distribution function for our test system.

\subsubsection{Calculation with $M=1$: \\ Two-body mean-field approximation.}

It is easy to see  that, when $M=1$, our QHNC equations
transform into the classical HNC equations. This happens  because all the
un-equal time  functions become zero and the equations take  a simple
form where only equal time  $hh$ functions survive.

\begin{figure}[H]
\vskip 0.3 in
\begin{center}
\includegraphics[scale=0.22]{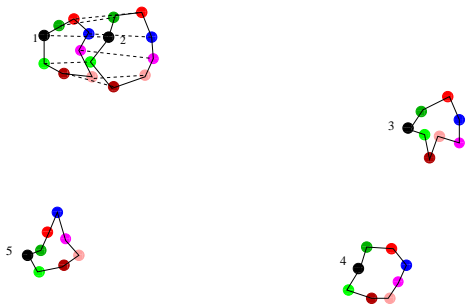}
\caption{ \label{fig:effective} Illustration of 
an effective interaction picture which emerges
at low density and in the high temperature limit using the 
ring-polymer isomorphism.
The color denotes the various positions of the atoms at different time-slices. 
The solid lines denote the $L$-lines, i.e., the spring-like interactions in the classical ring-polymer picture of the many-body path integral.
The dashed-lines denote the interaction between each bead of the polymer
with the corresponding bead of another polymer at the same time-slice.
}
\end{center}
\end{figure}

While we still work within the PTA starting point approach, we will use
the effective potential in conjunction with the HNC. 
One might question that such an effective-HNC approximation can be 
appropriate for  a quantum problem. 
First, note that the  effective potential (see discussion in 
Sec.~\ref{sec:low-density}) takes into account the effect of
the presence  of world-lines   in a way which  is exact in
the  zero-density  limit.  Therefore, it  should be  expected  to  be  a  good
approximation in the low-density regime also.  Furthermore, in the 
high-temperature limit we obtain the classical partition function and,
as a result, the HNC approximation is the appropriate treatment and is known to 
be accurate for a wide density range.

Our $M=1$ approximation should be viewed as 
a {\it two-body mean-field theory}, which is a consistent theory, 
and it should not be considered as just
the simple one time-slice approximation.
Namely, the physical meaning of the approximation is that
we account for all the quantum effects at the two-body level
exactly, and this defines
an effective integration which can be treated with classical statistical
mechanics. In Fig.~\ref{fig:effective} we illustrate the nature of the
approximation. At low density the most probable configurations
are the ones with mostly monomer configurations (ring-polymers with distance from
the other ring-polymers beyond the interaction range). The next most probable
occurrences are the clusters of two ring-polymers. A  treatment in which only
the ring-monomers are considered leads to the non-interacting ideal gas
approximation. The next level is to consider the two-body clusters and to solve
the two-ring polymer problem exactly by summing up all interaction terms 
using an infinite number of times-slices. 
This solution defines the  
effective interaction discussed in Sec.~\ref{sec:low-density},
which is then used to take into account the interaction of the center of mass
of these ring-polymers.

Notice that in the low density limit this treatment should 
be a good approximation. A second condition for this approach to be
accurate is that
the amplitude of the quantum-thermal fluctuations of the particle positions, 
as captured
by the de Broglie wavelength $\lambda_{\hbar\beta}$, should be small 
as compared to $r_s$ ($(4/3)\pi r_s^3 \rho=1$), i.e., 
\begin{eqnarray}
r_s >> \frac{\hbar}{\sqrt{2 \pi m k_B T}}.
\label{eq:debroglie}
\end{eqnarray}
 This is the analogue to the Born-Oppenheimer approximation which separates
the fast from the slow degrees of freedom. Here, we integrate out the 
degrees of freedom at all intermediate time-slices for each pair of
atoms. Therefore, a third condition for the applicability of this approach is 
that the time-scale for a third particle to 
approach an interacting pair and influence it, 
should be much larger than the imaginary-time
$\theta = \hbar \beta$. This implies that $r_s/v_{th} >> \theta$,
where $v_{th}$ is the classical atomic thermal velocity, and this leads to
the condition:
\begin{eqnarray}
r_s >> \frac{\hbar}{\sqrt{3 m k_B T}}.
\label{eq:thermal}
\end{eqnarray}
Notice that Eq.~\ref{eq:thermal} and Eq.~\ref{eq:debroglie} are compatible
with each other, giving approximately the same density-temperature 
boundary for the validity of this approximation. Notice that this condition is
very well satisfied for the case of liquid $^4$He at temperature of
the order of $T/\epsilon \sim 1$.

\begin{figure}[H]
\vskip 0.2 in
    \begin{center}
        \subfigure[]{
           \includegraphics[scale=0.32]{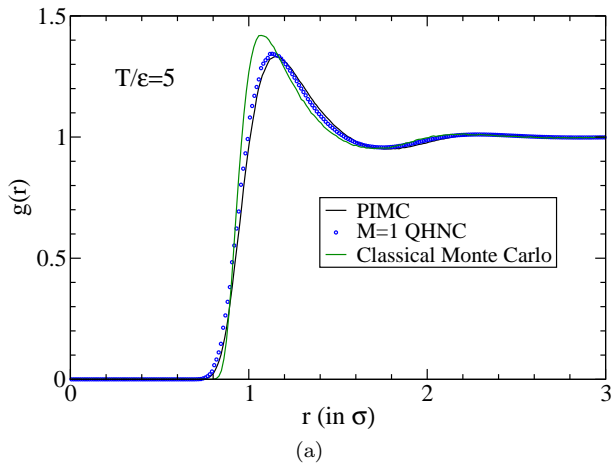}
           \label{fig:Temp5-m1}
        }\\
\vskip 0.4 in
        \subfigure[]{
            \includegraphics[scale=0.45]{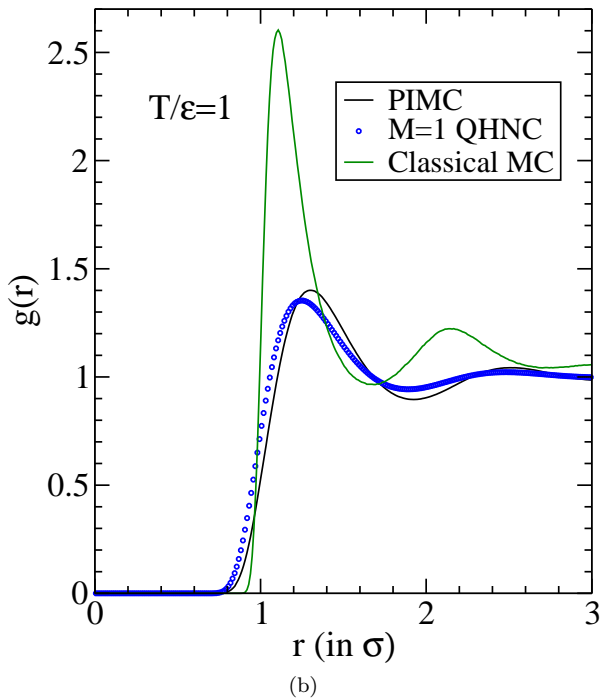}
            \label{fig:Temp1-m1}
        }
    \end{center}
    \caption{Comparison of $g(r)$ for particles interacting with Lennard-Jones
      potential for density $\rho=0.365 \sigma^{-3}$ and
      for temperature (a) $T/\epsilon=5$ and (b) $T/\epsilon=1$. The
  black  solid-line is  the result  obtained from  our PIMC simulation extrapolated to $M\to \infty$ and $N\to \infty$. 
  The blue circles represent   the result
  obtained  by  using  the   QHNC  equations  for $M=1$. }
\end{figure}

In an approach where we begin from the PTA as starting point, 
it appears that when we use the effective $h$-line in the QHNC equations
it produces spurious  diagrams in which a particle has
two world-lines.  However, these diagrams require a third particle
to be in close proximity with the two external particles.
However, at low density, the probability for that to occur is low.
At high and moderately low temperature, this can be also justified 
and a detailed discussion has  been provided  in Appendix~\ref{app:eff_int}.  

In order to test the accuracy  of this  scheme we  have applied it to the 
case of 
$^4$He as described by the Lennard-Jones interaction.  Results have  been obtained  for a rather high-density using $\rho = 0.365 \sigma^{-3}$ 
(the zero temperature liquid $^4$He equilibrium density). 
The matrix squaring method has  been used  to obtain the effective
$h$-line at  two different temperatures  namely $T=5$  and  $T=1$  (see Fig.~\ref{fig:Temp5-m1}  and Fig.~\ref{fig:Temp1-m1} for results). 
Notice that the results are reasonably close
to those obtained by PIMC using large enough number of time-slices
to achieve convergence.
We note that the  agreement between the $M=1$ 
approximation and the exact results  improves  as the  density  is
decreased or the temperature is increased.

\subsubsection{Calculation with $M=2$}

Within the PTA approach, i.e.,  starting from the expression of the path-integral with
the bare interaction, there is an inconsistency in replacing
the bare interaction with the effective interaction for $M>1$.
This does not present a conceptual problem when one starts
with the ETBDM approach\cite{Elba} because this interaction is our starting point.
However, surprisingly, as illustrated
in Appendix~\ref{sec:convergence}, while the use of the effective interaction with $M=1$ is exact at zero-density
and close to the exact at low-density, when the value of $M$
is increased, the effective interaction approach yields results 
which at first diverge from the exact solution at low $\rho$ and it takes high values of $M$ to achieve convergence.
Therefore, since we need to achieve convergence with respect to the number of
time slices $M$,  the PTA approach is superior to the the ETBDM approach. This is 
demonstrated in Appendix~\ref{sec:convergence}.

\begin{figure}
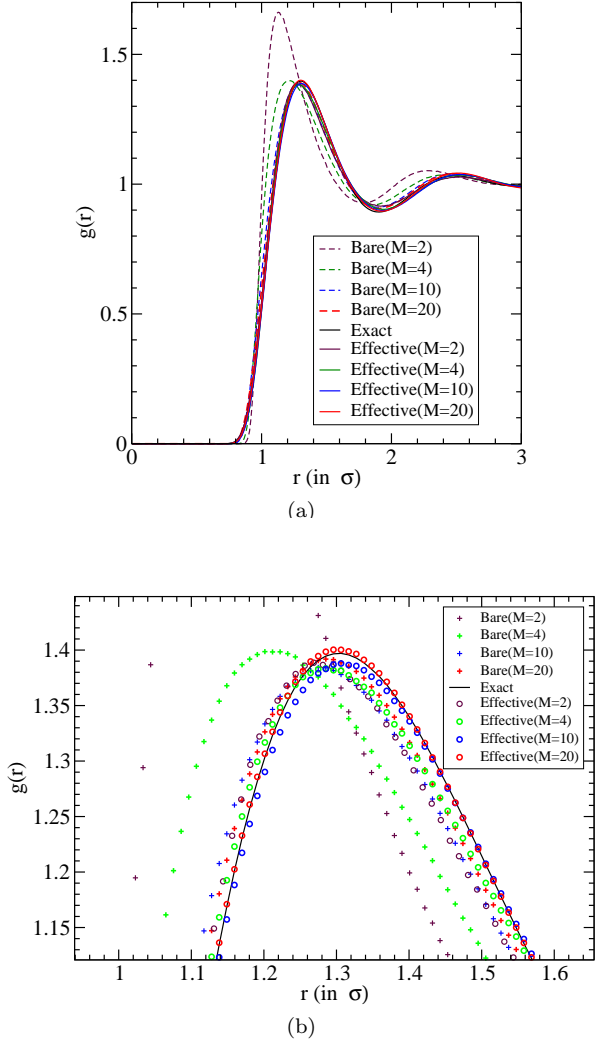

    \begin{center}
        \subfigure[]{
           \includegraphics[scale=0.32]{Fig19a}
           \label{fig:compare-all-m}
}\\
\vskip 0.3 in
        \subfigure[]{
           \includegraphics[scale=0.32]{Fig19b}
           \label{fig:compare-all-m-b}
}
    \end{center}
\caption{\label{fig:convergence-m}
(a) Convergence of $g(r)$ with the number of time-slices $M$ 
obtained at density of 0.365 $\sigma^{-3}$ starting from the bare 
interaction and with effective interaction as
a starting point and carrying a PIMC simulation. 
(b) Same as in (a) by expanding the scale near the
$g(r)$ maximum for clarity.}
\vskip 0.2 in
\end{figure}

On the contrary at densities of the order of $^4$He equilibrium density,
i.e., at density 0.365 $\sigma^{-3}$,  the $g(r)$ obtained for $M=2$ using the effective interaction in the PIMC simulation 
is close to that corresponding to the
$M \to \infty$ limit. Fig.~\ref{fig:compare-all-m} (Fig.~\ref{fig:compare-all-m-b}
for more detail near the main peak of $g(r)$)  
shows our results obtained with the PIMC method as a function of 
the number of time-slices $M$ starting from either the bare interaction or the effective interaction.
Notice that in fact the results for $M=2$ when using the effective interaction
are not too far from $M=\infty$. On the contrary
when starting from the PTA we need to use $M=20$ to achieve the same level of convergence.
This also allows us to
work with a few time slices without compromising the  accuracy of the
results for such densities.
Furthermore, as shown  in Sec.~\ref{sec:low-density},  when we  calculated
the first  order correction to $g(r)$  with just two time  slices but
using the effective interaction, the  results were reasonably good for
at least the high temperature  case.

\begin{figure}[htp]
\vskip 0.3 in
\includegraphics[scale=0.32]{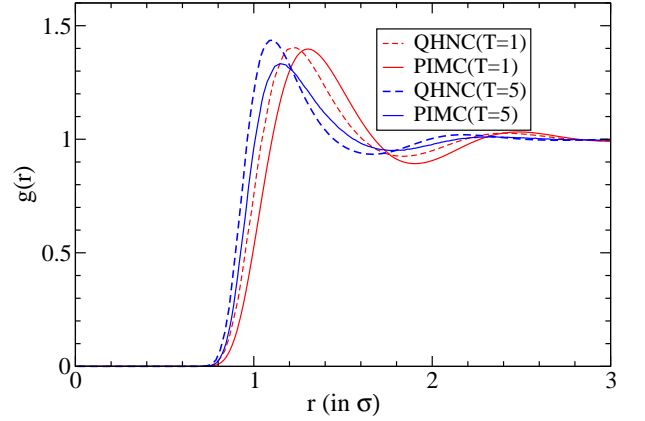}
\caption{ \label{fig:QHNC2}  Comparison of $g(r)$ between PIMC (solid  
line) and the QHNC (dashed line) with $M=2$
for temperature  $T/\epsilon =1$ (red) and  $T/\epsilon =5$ (blue). }
\vskip 0.3 in
\end{figure}

In Fig.~\ref{fig:QHNC2} we compare the results for $g(r)$ obtained by 
using the effective potential obtained after matrix squaring in conjunction 
with our QHNC equations with  $M=2$. Notice that the results of using QHNC are in some disagreement with the results of PIMC.

We would  like to emphasize that our QHNC equations  fail to include 
ladder diagrams  and the other diagrams that would be produced
by involving such ladder diagrams in Eq.~\ref{eq:qhnc}.  
Such diagrams which are of zeroth order were included by using the 
effective potential as discussed in the previous subsection for $M=1$.
In addition, the QHNC equations miss ladder diagrams where the ladder is formed by
connecting via $h$-lines parts of world-lines of different particles to construct the steps of the ladder (spatial-ladders). 
Including the exact contribution of such ladder diagrams is not 
an easy task.

Here, we outline a modification of the QHNC approach which allows
us to include an approximation of all such ladder diagrams.
In the regime where the contribution of such diagrams is small, 
a good estimate of their contribution could provide an accurate enough
calculation.

1) We  begin by using  the QHNC  equations to obtain  the pair
distribution  function  using the effective
potential for $\delta \tau = \hbar \beta/2$ (obtained by using the  matrix squaring method) because
we are dealing with $M=2$. 

2) Next, we approximate these missing 
diagrams. 

First, imagine that we have
summed up all the diagrams contributing to $g(r)$ (with the exception of the 
elementary diagrams) and the expression is given by the functional 
$g(\{L\},\{h\})$. Namely,
we would like to consider the sum of all the diagrams 
contributing to  $g$ as 
a functional of the $L$-lines and the $h$-lines  
for a reason that will become clear below.
To be consistent, the solution to the QHNC equations will be 
given by $g_Q(\{L\},\{h\})$, where $h$ is the $h$-line which 
corresponds to the effective potential.
Then, the contribution of the missing diagrams
is given by
\begin{eqnarray}
\Delta g =g(\{L\},\{h\})-g_{Q}(\{L\},\{h\}).
\label{eqn:deltag1}
\end{eqnarray}

Since the $\Delta g$ shown in Fig.~\ref{fig:QHNC2} is small compared to $g(r)$ itself
we will approximate $\Delta g$, i.e., the contribution of the missing diagrams
by using their simplified expressions at high-temperature.  
As we have discussed previously,  when $\delta \tau$ is small,  the $L$-lines
are Gaussian approximations to the delta-function, and in the limit of $\delta \tau \to 0$
they are true delta-functions leading to the classical limit, by ``collapsing''
the world lines in any given diagram. 
Therefore, a high-temperature approximation of these missing diagrams is given
by
\begin{eqnarray}
\Delta g =g(\{\delta\},\{h\})-g_{Q}(\{\delta\},\{h\}).
\label{eqn:deltag2}
\end{eqnarray}
$g(\{\delta\},\{h\})$ and $g_{Q}(\{\delta\},\{h\})$ are, respectively, 
the expression of the
exact distribution function and that obtained by solving the QHNC equations by  replacing the $L$-lines, with a delta-function.

\begin{figure}
    \begin{center}
        \subfigure[]{
           \includegraphics[scale=0.32]{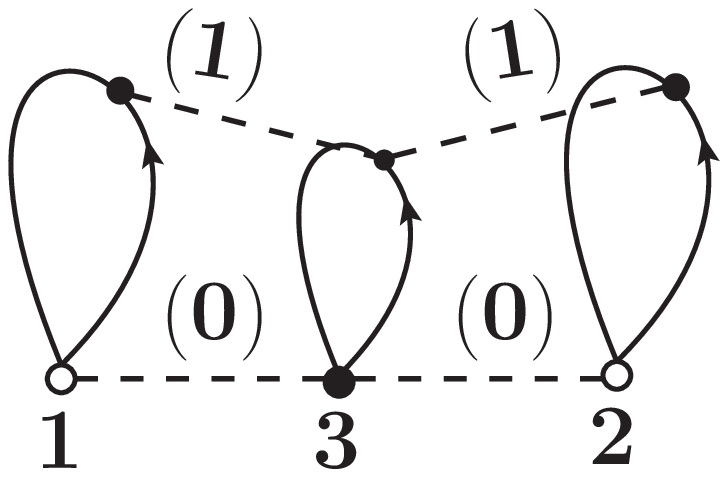}
           \label{fig:laddera}
        }
\hskip 0.6 in
        \subfigure[]{
            \includegraphics[scale=0.32]{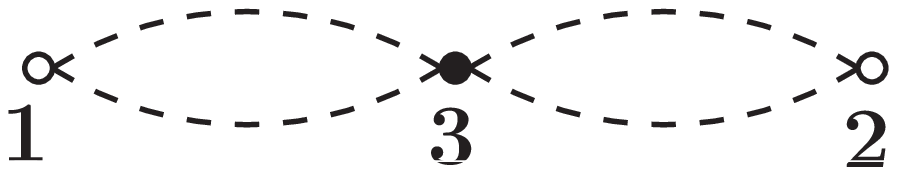}
            \label{fig:ladderb}
        }\\
\vskip 0.2 in
        \subfigure[]{
            \includegraphics[scale=0.32]{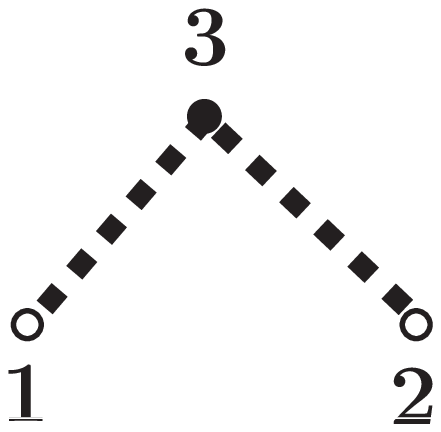}
            \label{fig:bold-line}
        }
\hskip 0.6 in
        \subfigure[]{
            \includegraphics[scale=0.32]{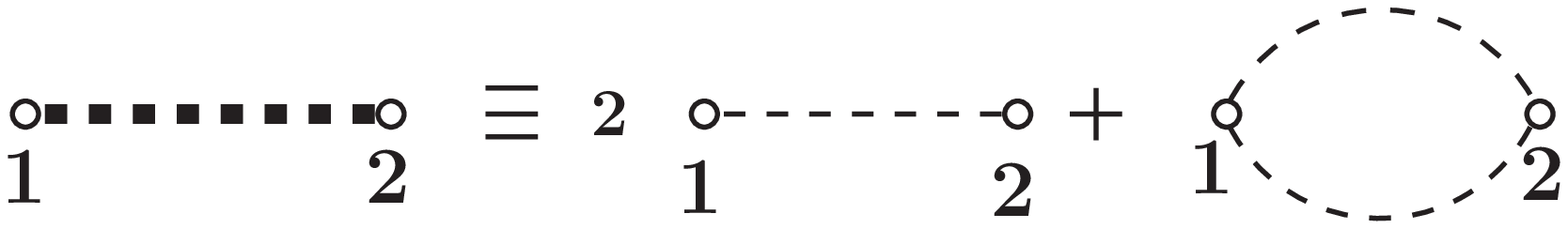}
            \label{fig:bold-exp}
        }
    \end{center}
    \caption{(a) A ladder diagram contributing to full distribution function
$g$ which is not obtained by the QHNC equations.  In the high-temperature
limit the diagram (a) reduces to diagram (b) because of the ``collapse'' of the
world-lines. This diagram is contained in the classical HNC diagram shown in 
(c) in which the bold-dashed line is expressed in terms of the
$h$-lines in the series illustrated in (d).}
\label{fig:ladder}
\end{figure}

However, as we have shown in Sec.~\ref{sec:low-density} 
$g(\{\delta\},\{h\})=g_{cl}$; namely
if we start from the sum of all the diagrams and we replace the $L$-lines
by a $\delta$-function, we obtain the classical limit.
As a result the contribution of the missing ladder diagrams from the QHNC equations can be approximated by
\begin{eqnarray}
\Delta g = g_{cl}-g_{Q}(\{\delta\},\{h\}).
\label{eqn:deltag}
\end{eqnarray}
Here, $g_{cl}$ is the pair distribution function obtained by using the
classical HNC. The fact that the HNC contains 
the high temperature approximation of the missing 
ladder diagrams is demonstrated diagrammatically in Fig.~\ref{fig:ladder}.
The reader  can understand the above statement by
studying Fig.~\ref{fig:ladder} (and its caption) in conjunction with 
our discussion of Sec.~\ref{sec:low-density}.

\begin{figure}[H]
\vskip 0.2 in
    \begin{center}
        \subfigure[]{
           \includegraphics[scale=0.32]{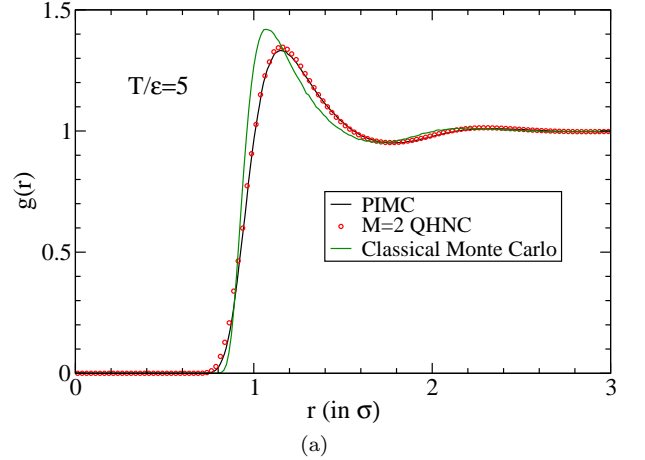}
           \label{fig:Temp5-m2}
        }\\
\vskip 0.3 in
        \subfigure[]{
            \includegraphics[scale=0.45]{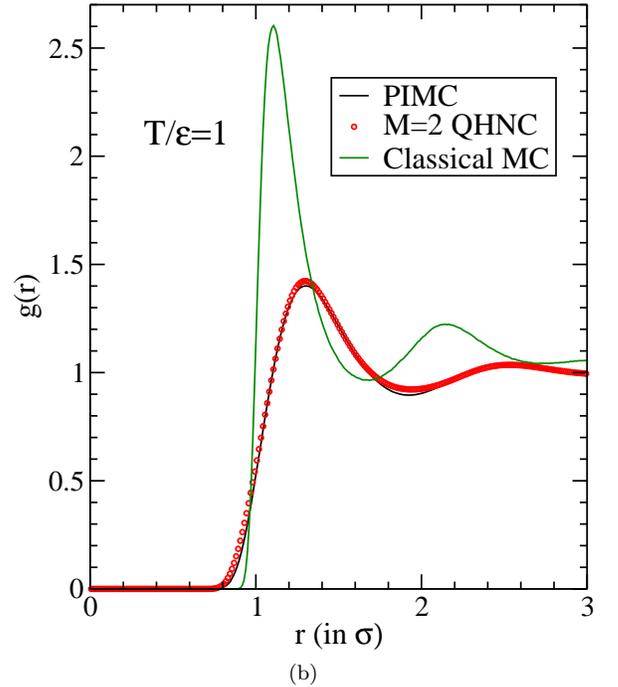}
            \label{fig:Temp1-m2}
        }
    \end{center}
    \caption{Comparison of $g(r)$ 
obtained by adding the 
  high-temperature approximation $\delta g$ to the  QHNC results 
for $M=2$ (red circles) for
  density $\rho=0.365 \sigma^{-3}$ and
      for temperature (a) $T/\epsilon=5$ and (b) $T/\epsilon=1$
to the PIMC simulation extrapolated to $M\to \infty$ and $N\to \infty$ (black-solid line).}
\vskip 0.2 in
\end{figure}

 We  worked at  density 0.365$\sigma^{-3}$ using two values of temperature,
$T/\epsilon=1$  and $T/\epsilon=5$, and the calculation of $g(r)$ is implemented
as follows.   For example, when using
$T/\epsilon=5$, we first obtained the $h_{eff}(T)$ at $T/\epsilon=10$ 
and we used it in the QHNC equations
to  obtain $g_{Q}$.  Then,  we used  this  $h_{eff}(10)$ to  obtain
$g_{Q}(\{\delta\},\{h\})$. The effective $h$ given by 
\begin{eqnarray}
h_{eff}={(h_{eff}(10)+1)}^2-1,
\end{eqnarray}
was used in the classical HNC equations to obtain $g_{cl}$.
Fig.~\ref{fig:Temp5-m2} and Fig.~\ref{fig:Temp1-m2} illustrate our
results for $T/\epsilon =5$ and $T/\epsilon =1$ and compares
them to the results obtained from PIMC simulation for 
$M \to \infty$. 
We  can clearly  see in  Fig.~\ref{fig:Temp5-m2} that the results from  QHNC
with $M=2$ 
(with the correction of $\Delta g$ described above given by Eq.~\ref{eqn:deltag}) 
agree with the  exact PIMC simulation.
Fig.~\ref{fig:Temp1-m2}  for $T/\epsilon=1$  indicates that 
the QHNC  equations with
$M=2$ (with the correction of $\Delta g$ given by Eq.~\ref{eqn:deltag}) 
 yields
more accurate  than the QHNC with $M=1$.
 
\subsubsection{Calculation with larger values of $M$}

If we apply the QHNC approach for $M>2$ we find that the results do 
not agree very well with the PIMC simulation. 
The QHNC equations fail to include diagrams in which world-lines involving more than one instant of
imaginary time where an interaction $h$-line occurs. These are diagrams which appear as ladders in 
the time direction (time-ladders). These diagrams begin to occur for $M>2$. 
In addition, the QHNC equations miss ladder diagrams where the ladder is 
formed by connecting parts of world-lines of different particles to construct the steps of the ladder 
(spatial-ladders). 

There are ways to generalize the QHNC equations to include such diagrams 
similar in spirit to those techniques used in conventional many-body perturbation theory\cite{BSE}
and this is one of the directions of our future work.

\section{Conclusions}
\label{conclusions}
We  have revived the old well-known method  of  cluster-expansion in classical statistical mechanics\cite{Mayer}  
by extending its application to the  many-body  path
integral and  we have derived a  diagrammatic expansion  
for the   pair   distribution  function.   
The
diagrammatic expansion  for the  pair distribution  function contains
both  connected  and  some factorizable/disconnected diagrams.  These disconnected
diagrams which do not cancel exactly involve disconnected parts with
common particles and can be paired with co-contributing factorizable
diagrams or some non-factorizable connected diagrams 
contributing with an opposite sign.  Because of this ``pairing'', 
these paired  
diagrams almost cancel and it can be shown that the cancellation is exact in the high-temperature limit.
By ignoring the contribution of these diagrams we can derive a cluster
expansion of connected diagrams which 
can be written as a formal power series expansion in the
particle density $\rho$. The series can be also thought of as 
an expansion where we keep all diagrams involving up to $n$-body clusters.
We compared our results to 
results we obtained by applying the PIMC technique which is exact
for distinguishable particles.
Surprisingly, the calculation of $g(r)$ including up to three-body diagrams 
(first order in 
$\rho$) gives results for the Lennard-Jones system which are accurate even 
for liquid-$^4$He equilibrium densities
and for temperature $T/\epsilon =1$ where $\epsilon$ is the well-depth of the Lennard-Jones potential.

We also generalized the hypernetted-chain approximation for our
quantum mechanical case (QHNC). We solved  the  QHNC equation for
the Lennard-Jones system for the case of one ($M=1$) and two ($M=2$)
imaginary-time slices
  along  with an effective  potential obtained  from the
matrix squaring technique. 
This was found to be accurate down to moderately low temperature  
($T/\epsilon=1$). We find that our QHNC equations need to be generalized
to tackle the case of $M > 2$.

Our method is generalizable to the case of identical particles
and in particular to the case of fermions.
We hope that this method can provide useful results to 
compare with the results obtained by applying 
quantum Monte Carlo (QMC) methods in this latter case where
the QMC results are not exact. We hope to provide results for
this case in the near future.

\vskip 0.5 in

\section{Acknowledgments}

This work was supported in part by the U.S. National High Magnetic Field 
Laboratory, which is
funded by NSF DMR-1157490 and the State of Florida.

%


\appendix

\section{\label{app:eff_int} Spurious diagrams}

\begin{figure}[htp]
\vskip 0.3 in
\begin{center}
\includegraphics[scale=0.30]{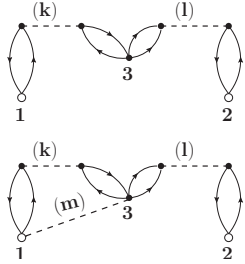}
\caption{Examples of spurious terms that appear in effective HNC expansion in which particle 3 appears to have two world-lines. \label{fig:spurious1}}
\end{center}
\vskip 0.3 in
\end{figure}

In the effective-HNC approximation (i.e., our $M=1$ approximation) the terms up to zeroth order match exactly with the exact quantum expansion up to zeroth order. However, for terms which are higher than zeroth order, the effective-HNC approximation is not able to account for all the terms that appear in the exact quantum expansion. There are spurious terms which appear in the expansion in which the particles have two world-lines (see Fig.~\ref{fig:spurious1}).

\begin{figure}[htp]
\vskip 0.3 in
\begin{center}
       \subfigure[]{
           \includegraphics[scale=0.32]{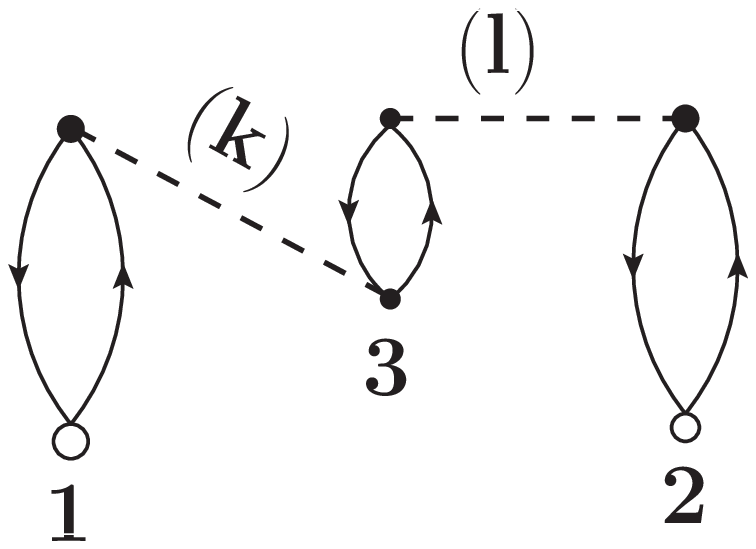}
    
        }
\hskip 0.6 in
        \subfigure[]{
            \includegraphics[scale=0.32]{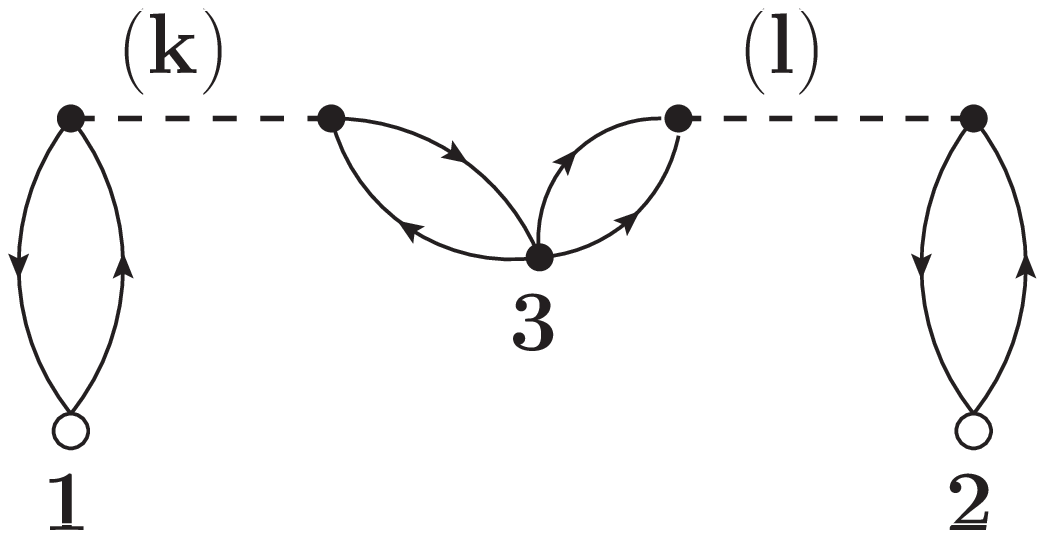}
            
        }
\caption{(a) A term that appears in the exact quantum expansion. (b)The spurious counterpart of (a) in which particle labeled 3 appears to have two world-lines. \label{fig:spurious2}}
\end{center}
\vskip 0.3 in
\end{figure}

\begin{figure}[htp]
\vskip 0.3 in
\begin{center}
\includegraphics[scale=0.35]{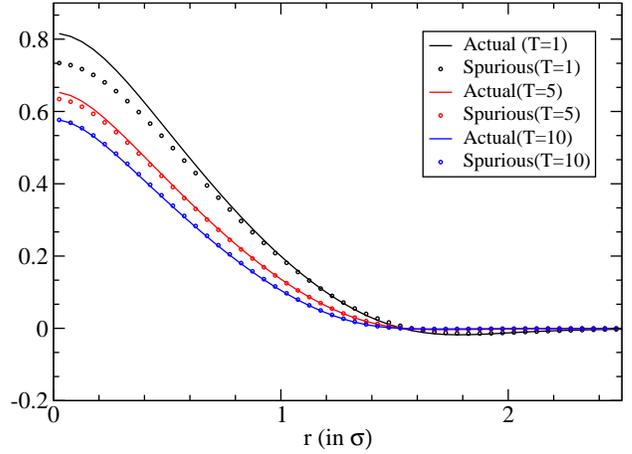}
\caption{Comparison of the exact term and its spurious counterpart for the case where $k=l=1$. The total number of division used is 16 ($M=16$) 
\label{fig:all1}}
\end{center}
\vskip 0.3 in
\end{figure}

\begin{figure}[htp]
\vskip 0.3 in
\begin{center}
\includegraphics[scale=0.35]{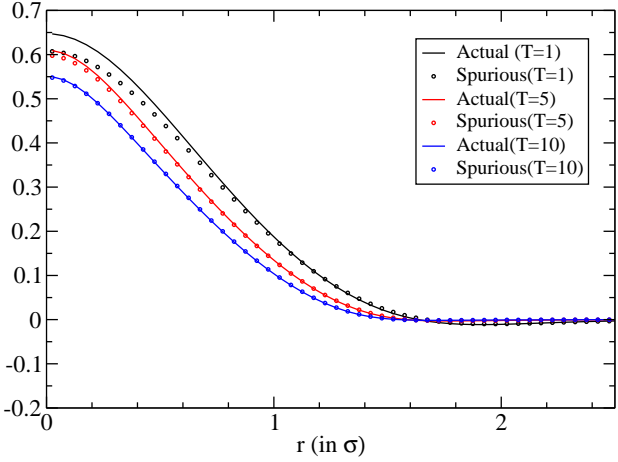}
\caption{Comparison of the contributions of the exact diagram and that of 
its spurious counterpart for the case where $k=1$ and $l=4$ for $M=16$. \label{fig:all2}}
\end{center}
\vskip 0.3 in
\end{figure}

The high-temperature limit of the effective $h$-line is the classical $h$-line. Also, both the exact quantum pair distribution function and the pair distribution function obtained from effective-HNC become the classical pair distribution function in the high-temperature limit. For any spurious diagram that appears in the effective-HNC there is a unique counterpart diagram in the exact expression for $g(r)$ (see Fig.~\ref{fig:spurious2} as an example).
Any such spurious diagram and its counterpart in the actual expansion 
have approximately the same contribution in certain temperature regime (as temperature becomes high the two match exactly). We have compared the 
two terms (exact and its spurious counterpart) that appear in Fig.~\ref{fig:spurious2} for different temperature to see how close these two terms are down to
temperature $T/\epsilon =1$. 
We kept the number of time slices fixed at 16 and have 
considered the case where $k=l=1$ and the case where $k=1$ and $l=4$. The results are shown in  Fig.~\ref{fig:all1}, and Fig.~\ref{fig:all2}.
Since the expansion parameter is $\rho$ we expect the 
effective-HNC scheme to be more accurate in the low density and high temperature regime.

\section{\label{sec:convergence} Convergence with respect to the number of time-slices $M$}

\begin{figure}[H]
\vskip 0.2 in
    \begin{center}
        \subfigure[]{
           \includegraphics[scale=0.32]{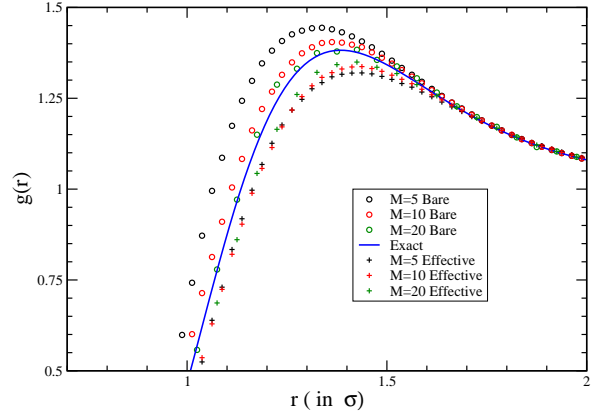}
           \label{fig:compare-bare-effective}
        }
\vskip 0.2 in
        \subfigure[]{
            \includegraphics[scale=0.32]{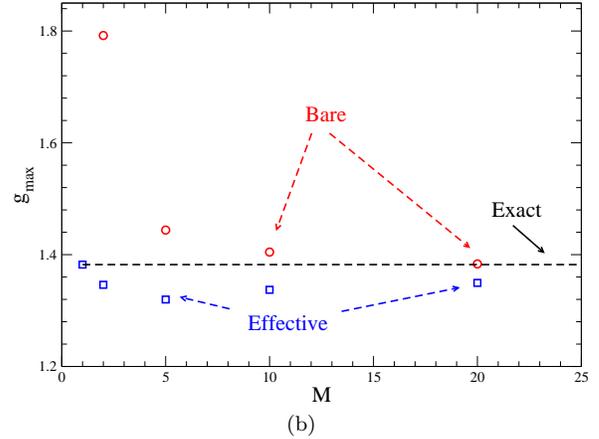}
            \label{fig:compare-bare-effective-max}
        }
    \end{center}
    \caption{(a) \label{fig:compare-convergence}
 Convergence of $g(r)$ with the number of time-slices $M$ 
obtained at zero-density starting from the bare interaction, i.e., within the
PTA approximation and with the ETBDM (effective interaction) as
a starting point. (b) Convergence of the maximum of $g(r)$ as a function
of $M$ with the two approaches.}
\vskip 0.1 in
\end{figure}

The approach of using the ETBDM in PIMC simulations and in our approach
 has the following problem for low density illustrated in Fig.~\ref{fig:compare-convergence} 
for zero density. Fig.~\ref{fig:compare-convergence} presents the results obtained for $g(r)$
for various values of $M$ using (a)
 the bare interaction and (b) the effective interaction corresponding to
the particular value of $M$. It can be clearly seen that when we use
the bare interaction the results for $M=20$ are very close to
the exact results. When, however, we use the effective interaction,  even though the $g(r)$ for $M=1$
is already the exact $g(r)$, the results for $M=2,3,4,5$  move  
further away from exact solution and, for larger values of $M$, they begin moving closer
to the exact solution. However, even for $M=20$ they are still far from the
exact results. Therefore, the convergence is faster when the bare interaction
is used.  Thus, if one did not know that the starting $g(r)$, which 
corresponds to $M=1$ was the exact solution, and one uses the convergence 
criterion to extract the correct $g(r)$ it is better to use the bare interaction,
i.e., the approach based on the PTA.
 This creates a problem for the convergence with $M$ in the PIMC approach at low density when
starting from the ETBDM.  While this seems surprising at first, it can be easily understood
within our approach. While the effective potential for
$M=1$ corresponds to the exact $g(r)$ at
zero density (and presumably a good approximation at low density), when dealing with $M>1$
using the effective potential corresponding to inverse temperature $\beta/M$ for the $M$ intermediate 
time-slices leads to the following problem. 
The density sub-matrix needed to connect two consecutive time-slices separated by $\delta\tau = \hbar \beta/M$ 
is the full density matrix including its off-diagonal matrix elements in order to connect the points at $\vec r_1^{(M)},\vec r_2^{(M)}$ with the points $\vec r_1^{(M+1)},\vec r_2^{(M+1)}$. 
Instead, only the diagonal matrix elements are utilized when we use 
the effective potential. This yields the correct $g(r)$ when $M$ is infinite and in that limit 
the bare interaction is good enough. 
This leads to the results presented in Fig.~~\ref{fig:compare-convergence}
where, if we use the effective interaction, the density matrix is already the exact at the $M=1$ level, 
and by trying to go beyond that by increasing the value of $M$, using the effective interaction at $M>1$, the results first 
diverge from the exact results and, then, slowly approach the same 
solution after very large value of $M$ (see Fig.~\ref{fig:compare-bare-effective-max}).
 
On the contrary at densities of the order of $^4$He equilibrium density,
i.e., at density 0.365 $\sigma^{-3}$,   the convergence when using 
the effective interaction is much better than when using the bare interaction as
starting point. 
This is illustrated in Fig.~\ref{fig:compare-all-m-b}, which  is the same as Fig.~\ref{fig:compare-all-m} 
but we have expanded the region near the peak.
We can now clearly see that for this high density using the effective
interaction we achieve much faster convergence with $M$ as compared to using
the bare interaction.

\end{document}